\documentclass[preprints,article,accept,moreauthors,pdftex]{Definitions/mdpi} 

\usepackage{graphics} 
\usepackage{epsfig} 
\usepackage{mathptmx} 
\usepackage{times} 
\usepackage{amsmath} 
\usepackage{amssymb}  
\usepackage{textcomp}
\usepackage{multirow}
\usepackage{colortbl}
\usepackage{xcolor}
\usepackage{hhline}
\usepackage{bm}
\usepackage{hyperref}
\usepackage{lipsum}
\usepackage{mathtools}
\usepackage{cuted}
\usepackage{todonotes}
\usepackage{subfig}
\usepackage{comment}

\graphicspath{{figs/}{evaluation/}}

\newcommand{\newcolor}[1]{\color{black}}
\definecolor{red}{cmyk}{0,0.89,0.94,0.28}

\firstpage{1} 
\pubvolume{xx}
\issuenum{1}
\articlenumber{5}
\pubyear{2020}
\copyrightyear{2020}
\history{}

\pdfoutput=1

\Title{Cloud2Edge Elastic AI Framework for Prototyping and Deployment of AI Inference Engines in Autonomous Vehicles}

\newcommand{\orcidauthorSorin}{0000-0003-4763-5540}
\newcommand{\orcidauthorTibi}{0000-0003-4311-0018}
\newcommand{\orcidauthorBogdan}{0000-0001-6169-1181}
\newcommand{\orcidauthorAndrea}{0000-0002-5048-8070}
\newcommand{\orcidauthorFederico}{0000-0001-6463-8722} 
\newcommand{\orcidauthorLeonardo}{0000-0003-2886-8445}

\Author{Sorin~Grigorescu $^{1}$\orcidauthorSorin{}, Tiberiu~Cocias $^{1}$\orcidauthorTibi{}, Bogdan~Trasnea $^{1}$\orcidauthorBogdan{}, Andrea~Margheri $^{2}$\orcidauthorAndrea{}, Federico~Lombardi $^{2}$\orcidauthorFederico{}* and Leonardo~Aniello $^{2}$\orcidauthorLeonardo{}}

\AuthorNames{Sorin~Grigorescu, Tiberiu~Cocias, Bogdan~Trasnea, Andrea~Margheri, Federico~Lombardi, Leonardo~Aniello}

\address{%
$^{1}$ \quad  Robotics, Vision and Control Laboratory (ROVIS) at Transilvania University of Brasov and Elektrobit Automotive; s.grigorescu@unitbv.ro, tiberiu.cocias@unitbv.ro, bogdan.trasnea@unitbv.ro\\
$^{2}$ \quad  School of Electronics and Computer Science at the University of Southampton; a.margheri@soton.ac.uk, f.lombardi@soton.ac.uk, l.aniello@soton.ac.uk}

\corres{Correspondence: s.grigorescu@unitbv.ro}

\abstract{Self-driving cars and autonomous vehicles are revolutionizing the automotive sector, shaping the future of mobility altogether. Although the integration of novel technologies such as Artificial Intelligence (AI) and Cloud/Edge computing provides golden opportunities to improve autonomous driving applications, there is the need to modernize accordingly the whole prototyping and deployment cycle of AI components. This paper proposes a novel framework for developing so-called \textit{AI Inference Engines} for autonomous driving applications based on deep learning modules, where training tasks are deployed elastically over both Cloud and Edge resources, with the purpose of reducing the required network bandwidth, as well as mitigating privacy issues. Based on our proposed data driven V-Model, we introduce a simple yet elegant solution for the AI components development cycle, where prototyping takes place in the cloud according to the Software-in-the-Loop (SiL) paradigm, while deployment and evaluation on the target ECUs (Electronic Control Units) is performed as Hardware-in-the-Loop (HiL) testing. The effectiveness of the proposed framework is demonstrated using two real-world use-cases of AI inference engines for autonomous vehicles, that is \textit{environment perception} and \textit{most probable path prediction}.}

\keyword{Autonomous Vehicles, Self-driving Cars, Artificial Intelligence, Deep Learning, Cloud Computing, Edge Computing.}

\begin{document}



\section{Introduction}
\label{sec:introduction}



Among the new technological trends arisen in the last decade, \textit{Autonomous Driving} has gained a lot of attention, with significant effort and resources invested by both academia and enterprises. The breakthroughs in self-driving cars has been made possible by the emergence of novel algorithms and real-time computing systems in the field of Artificial Intelligence (AI), Deep Learning (DL), Internet-of-Things (IoT) and Cloud computing~\cite{Grigorescu_JFR_2020}.

An Autonomous Vehicle (AV) is an intelligent agent which observes its environment, makes decisions and performs actions based on these decisions. Autonomous driving applications aim to enable vehicles to automatically control their behavior. Traditionally in classical cars, these aspects and functionalities are the exclusive responsibility of human drivers, including understanding of the driving scene, path prediction, behavioral planning and control. The development and deployment of accurate AI models which can automatize these tasks is fundamental to support the progress of autonomous driving. Addressing these challenges lies within the realms of \textit{Data Science}, with AI and DL proving to be successful tools to generate sufficiently accurate representations of real-world processes. Deep Learning (DL) and Deep Neural Networks (DNN) have become leading technologies in many domains, enabling vehicles to perceive their driving environment and take actions accordingly. Although several cloud-based solutions have been proposed for the automatic reconfiguration and adaptation of distributed platforms in Industry 4.0~\cite{Villalonga_2020}, automotive grade systems have been scarcely reported.

Nowadays, numerous software tools are available, providing solutions for each particular stage in the development of an AI-based application. A key benefit for a driving function engineer is to have a comprehensive framework for enabling both the prototyping of the AI components, as well as their deployment and evaluation on target Edge devices such as Electronic Control Units (ECUs). General AI tooling systems such as MATLAB~\cite{Matlab2018}, Amazon Web Services (AWS)~\cite{AVS2020} or AI-One~\cite{AIONE2020} provide functionalities to enable users to design and integrate custom AI modules. However, these tooling systems are neither tailored for the automotive sector, nor for the specificity of the autonomous driving applications. One of the major drawbacks is their lack of alignment to the automotive functional safety standard ISO26262~\cite{Salay2017}, particularly to ASIL (Automotive Safety Integrity Level) hierarchy. On the other hand, major automotive and chip manufacturing companies are developing their own in-house AI frameworks, such as Tesla's \textit{Full Self-Driving} and \textit{Dojo} computers~\footnote{Tesla's Dojo --- \url{https://syncedreview.com/2020/08/19/tesla-video-data-processing-supercomputer-dojo-building-a-4d-system-aiming-at-l5-self-driving/}}, NVDIA's \textit{DriveWorks} platform~\footnote{NVIDIA DriveWorks --- \url{https://developer.nvidia.com/drive/driveworks}}, or Uber's Michelangelo training framework~\footnote{Uber's Michelangelo --- \url{https://eng.uber.com/michelangelo-machine-learning-platform/}}. The problem with these tooling systems is that they are either designed to be used solely by the developing company (e.g. Tesla, Uber, etc.), or are build to operate only on specific hardware, as in the case of NVIDIA's AI solutions.

In this paper, we first analyze the development process of AI-based autonomous driving applications and identify its current limitations. Secondly, we propose the \emph{Coud2Edge Elastic EB-AI Framework} as a revised prototyping and deployment workflow for building AI based solutions for autonomous driving. We introduce the novel concept of \textit{AI Inference Engine}, which encompasses a simple yet elegant solution to the training, evaluation and deployment of AI-based components for self-driving cars. Throughout the paper, we refer to a real scenario, where autonomous driving applications are developed at Elektrobit (EB) Automotive~\footnote{Elektrobit Automotive --- \url{https://www.elektrobit.com/}}, a world-wide supplier of embedded and connected software products and services for the automotive industry.

\begin{figure}
	\centering
	\begin{center}
		\includegraphics[scale=0.47]{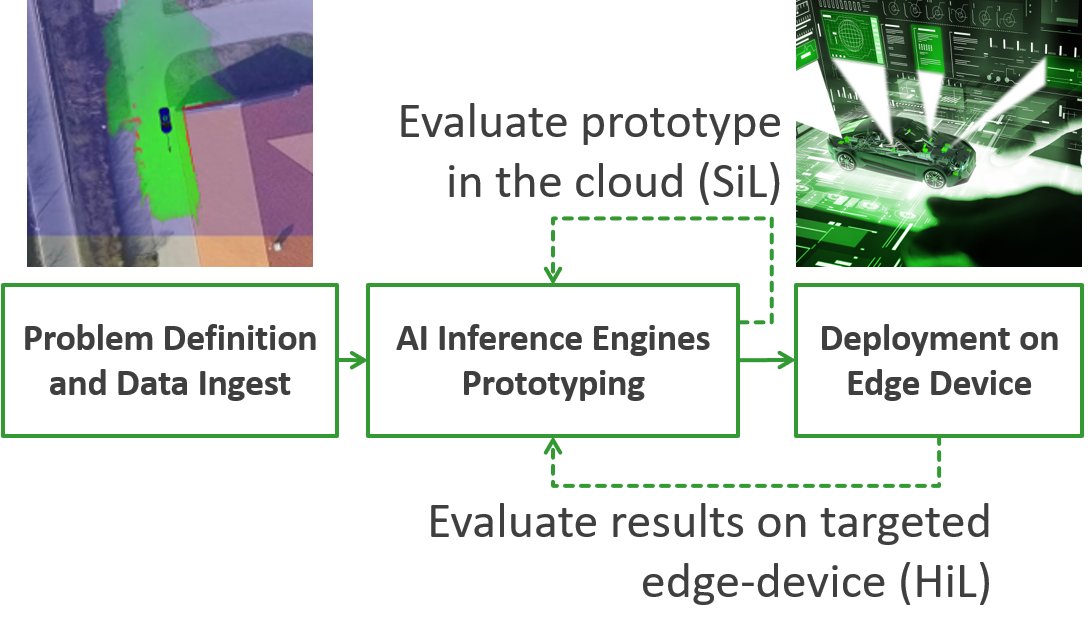}
		\caption{\textbf{AI Inference Engine concept}. Deep Learning based inference engine evaluated at prototyping level within the Cloud as Software-in-the-Loop (SiL), as well as on the targeted edge device ECU, as Hardware-in-the-Loop (HiL).}
		\label{fig:inference_engine_workflow}
	\end{center}
	\vspace{-2.2em}
\end{figure}

\subsection{The EB-AI Framework}

The key differences of the EB-AI framework provides with respect to tooling systems such as the ones listed above are:

\begin{itemize}
	\item ability to digest automotive specific input data such as video streams, LiDAR and/or Radar data;
	\item prototyping and development of AI Inference Engines in the Cloud using a SiL paradigm, agnostic of low-level AI libraries such as TensorFlow~\cite{abadi2016tensorflow}, PyTorch~\cite{paszke2017automatic}, or Caffe~\cite{Caffe};
	\item providing tunable state of the art DNN architectures for a broad spectrum of autonomous driving applications;
	\item ability to deploy and evaluate the obtained AI Inference Engines via HiL testing on edge devices (e.g. target ECUs).
\end{itemize}

The diagram in Fig.~\ref{fig:inference_engine_workflow} illustrates the development workflow of an AI Inference Engine based on the EB-AI framework, according to our own data driven V-Model detailed in Section~\ref{sec:architecture}. Once the specific problem to solve has been defined and enough data has been collected, that data is ingested at prototyping level by a DNN which stands at the core of the inference engine. During this stage, the DNN is trained, evaluated and refined within the Cloud according to the Software-in-the-Loop (SiL) principles. Finally, the inference engine is deployed on a target Edge device inside a vehicle and evaluated again in real-world scenarios as Hardware-in-the-Loop (HiL). This process allows us to refine the engine even further, according to a continuous feedback loop aimed to keep improving applications over their entire life span. The novelty of the AI Inference Engine concept lies in the effective integration of the DNN's training, evaluation as SiL and testing using the HiL paradigm.

A key aspect to consider in the EB-AI workflow is the huge amount of data required to train DNNs. This data can be either synthetically generated (first stage) or collected via real sensors mounted on test vehicles (last stage). In both cases, the data has to be made available at training stage. Although the training of DNNs can be highly parallelized~\cite{batiz2012parallelizing}, with unprecedented levels of parallelization reached by leveraging the recent wide availability of GPUs, it is still prohibitive to have enough computational power to process the amount of data required by autonomous driving applications. As a consequence, all Cloud computing providers offer commoditized AI solutions (e.g., IBM’s Watson AI, Amazon AI, Microsoft’s Azure AI and Google’s Cloud AI) for scalable computing training facilities. However, the following limitations hold at AI Inference Engine prototyping stage, when large quantities of data have to be uploaded to the Cloud: \textit{i}) data upload impracticality due to bandwidth bottlenecks and latency~\cite{Nangare2018}, \textit{ii}) difficult enforcement of privacy requirements (e.g. GDPR~\cite{Wiles2018}), since part of the collected raw data can be sensitive and thus cannot be shared with cloud providers.

\subsection{Contributions}

We propose a solution to overcome the presented limitations by exploiting recent developments in Over-The-Air (OTA) systems for AVs~\cite{greenough2016connected, khurram2016enhancing}, enabling the use of an AV as an additional computational device. These computing resources localized on the Edge ECUs, also referred to as \emph{Edge computing}, can be integrated with Cloud resources to provide an \textit{Elastic} distributed infrastructure in which the training can be collaboratively carried out~\cite{HuangMFLG17,8270639}. Furthermore, Edge devices can limit the data transmitted to the Cloud by pre-processing local raw data or pre-training partial models~\cite{8487352}, which in turn can also help mitigate privacy concerns, since raw data sharing can be directly avoided. This kind of hybrid deployment over the Cloud and Edge resources calls for a modular framework for AI-based autonomous driving applications. To this aim, the proposes EB-AI architecture provides a modular toolchain that enables the deployment of autonomous driving applications across Edge and Cloud resources.

Therefore, the main contributions of the paper are:

\begin{itemize}
	\item a simple yet elegant AI Inference Engine concept, based on the SiL and HiL principles;
	\item a data-driven V-Model approach guiding the design of AI-based autonomous driving applications;
	\item a modular Cloud2Edge AI framework for autonomous driving applications coined EB-AI;
	\item an Elastic framework able to overcome network bandwidth and privacy concerns by dynamically deploying deep learning training tasks among Edge and Cloud;
	\item the development of two real-world AI Inference Engines for \textit{environment perception} and \textit{most probable path prediction};
	\item a discussion on the advantages of such a hybrid deployment, in terms of training parallelization, privacy-preservation, fault tolerance and scalability.
\end{itemize}

\smallskip
\noindent
\emph{Structure of the Paper.} 
%
Section~\ref{sec:related_work} reports most relevant background concepts and related work. Section~\ref{sec:architecture} presents the V-Model approach and the proposed EB-AI framework. Section~\ref{sec:cloud_edge} discusses the deployment of the architecture across Cloud and Edge devices. Finally, two relevant AI Inference Engines are detailed in Section~\ref{sec:use_cases}, while the conclusions are stated in Section~\ref{sec:conclusions}.

	
\section{Background and Related Work}
	\label{sec:related_work}

In the following, we overview the main concepts of DL, report on general DL tooling systems and highlighting the main limitations of available tools and learning environments for automotive applications.

\subsection{Deep Learning Overview}

\textit{Deep Learning} is a branch of machine learning which leverages on large DNN architectures to build a layered representation of the input data~\cite{bengio2013representation}. DNNs are universal non-linear function approximators, built using multiple hidden layers, and can be classified based on their architecture. \textit{Convolutional Neural Networks} (CNNs) are mainly used for processing spatial information, such as images, and can be viewed as image feature extractors. Recurrent Neural Networks (RNN) are especially good in processing temporal sequence data, such as text, or video streams. Different from conventional neural networks, an RNN contains a time dependent feedback loop in its memory cell. Due to the high number of layers, DNNs are difficult to train using vanilla backpropagation algorithms. As an example, in the case of auto-encoders~\cite{hinton1994autoencoders}, each layer is trained separately, then backpropagation is used to train the whole network. To avoid overfitting (i.e. trained models that cannot generalize or predict unseen values), common methods such as Dropout~\cite{srivastava2014dropout} are used to regularize the training.

DNNs are effective in domains characterized by a high number of features, such as computer vision and natural language processing. Training of DNNs critically requires large annotated datasets which have been increasingly released in the last decade, especially in the computer vision field. One of the first large collection of annotated images is the ImageNet database~\cite{ImageNet_ILSVRC15}, which contains $1.5$ mil images representing $1000$ object classes in a variety of shapes, poses and illumination conditions. Although images relevant to autonomous driving systems are only a subset of the whole collection, the complete database were used to pre-train the first convolutional layers of object detectors used for driving scene perception~\cite{Grigorescu_JFR_2020}.

In the last years, mainly due to the increasing research interest in autonomous vehicles, many driving datasets were made public. These vary in size, sensory setup and data format, commonly comprising of synchronized ego-data, video, LiDAR, Radar, ultrasonic, inertial measurements and GPS datastreams. Among others, popular ones are the KITTI Vision Benchmark dataset (KITTI)~\cite{KITTI2013} by the Karlsruhe Institute of Technology (KIT), NuScenes~\cite{nuscenes2019} by Aptive, or Cityscapes~\cite{Cityscapes2018} by Daimler, Max Planck Institute and TU Darmstadt.

Since acquiring large training datasets is a demanding process, usually performed via manual annotation, alternative methods have been explored for synthetic data generation and simulation engines. In our previous work we have proposed and patented a semi-parametric approach to one-shot learning, coined \emph{Generative One-Shot Learning} (GOL)~\cite{grigorescu2018generative}. Given as input single one-shot objects (or generic patterns and templates), together with a small set of regularization samples, GOL  uses the sample to drive the generative process by outputting new synthetic data. Specifically, it permits to generalize on unseen data, while increasing the classification accuracy on synthetic data as much as possible.

\subsection{Deep Learning Libraries and Tools}

Some of the well-established low-level AI libraries are Tensorflow~\cite{abadi2016tensorflow} by Google, PyThorch~\cite{paszke2017automatic} by Facebook, the Cognitive Neural ToolKit (CNTK)\footnote{CNTK --- \url{https://docs.microsoft.com/en-us/cognitive-toolkit/}} by Microsoft, and Caffe2~\cite{Caffe} from Berkeley University. The APIs are mainly available for C/C++ and Python  languages, although other notable implementations exists, such as Weka~\cite{hall2009weka} which is written in Java. For facilitating their usage and large scale adoption, these libraries are integrated into high-level frameworks (e.g. Lasagne\footnote{Lasagne --- \url{http://lasagne.readthedocs.org/en/latest/}}).

For fast computation, training commonly takes place on GPUs interconnected as clusters, which are available as a service on the major Cloud platforms, e.g. Microsoft Azure and Amazon AWS. After training, the DNN models are downloaded into edge devices for inference. In this case, edge devices can be smartphones, PC commuters, or target ECU devices part of automotive applications.

In order to exploit parallel computation, the training can be distributed based on data partition (e.g. among threads, different machines,  GPUs, etc.) or  model partition~\cite{xing2016strategies}. Scalable deployments are enabled by distributed computing platforms such as Storm~\cite{toshniwal2014storm} and Spark~\cite{zaharia2010spark}. In particular, \mbox{Trident-ML}\footnote{Trident-ML --- \url{https://github.com/pmerienne/trident-ml}} is a real-time library for machine learning upon Storm, while CaffeOnSpark~\cite{noel2016large} and TensorSpark~\cite{tensorspark} are Spark integrations of Caffe and TensorFlow, respectively.

To overcome bandwidth and privacy issues when uploading data to the Cloud, federated learning~\cite{YangLCT19} offers a technical framework to scale learning horizontally or vertically across devices~\cite{KonecnyMRR16}. Edge computing is therefore exploited to carry out pre-training, whose results are integrated on the Cloud and then offloaded on the Edge for on-device inference~\cite{8681645}. 

The above mentioned frameworks are designed solely for training purposes within the Cloud. In comparison, the main advantage of EB-AI over them is our proposed Elastic approach, which considers both training and inference as a joint task distributed onto powerful GPU clusters residing in the Cloud and Edge computing devices available in each autonomous vehicle belonging to a connected fleet of cars. Additionally, as opposed to the more general AI frameworks presented in this subsection, EB-AI has been designed specifically for automotive purposes, following the constraints of Automotive SPICE and the ISO26262~\cite{Salay2017} functional safety standard, along with a proposed data driven V-Model for building AI Inference Engines.

\subsection{AI Frameworks in Autonomous Driving}

The application of deep learning, and more general of AI, to the automotive industry has grown significantly in the last few years. AI and Deep Learning are used to obtain human-like behaviors in automated driving~\cite{luckow2016deep}, such as environment perception and scene understanding (e.g. detecting traffic signs, traffic participants, road obstacles, etc.), or trajectory planning and control. Deploying AI Inference Engines inside autonomous vehicles requires overcoming limitations of platform dependencies and limited computation resources. Therefore, new transportation vehicles are equipped with embedded ECU devices specialized for AI and exploited to ensure adequate performance and energy efficiency~\cite{brilli2018convolutional}.

Overall, the increasing interest in AI Inference Engines for the automotive industry has been driven by the field of autonomous vehicles. While a number of AI-enabled functions are already deployed in cars and interacting with drivers~\cite{fridman2017autonomous}, it is predicted that full-fledged AVs, that is SAE levels 4 and 5~\cite{SAE}, will require significant additional cost for developing the new required AI solutions~\cite{litman2019autonomous}. As such, there is the need to optimize the design, deployment and maintenance of automotive AI Inference Engines by reducing their complexity and by improving accuracy via the access to (real-time) data. Lu et al. proposed a work close to our~\cite{234807}, where they introduced distributed learning based on self-driving cars as Edge computing devices. However, they only train local models on the cars and then aggregate the parameters of the neural network on a Parameter EdgeServer. Conversely, we propose a solution to split training tasks between cloud and edge, in order to deploy training task which process sensitive data only over the edge while deploying the other over both cloud and edge.

Car manufactures and their suppliers are both developing their own in-house AI frameworks. Example of such systems are Tesla's \textit{Full Self-Driving} and \textit{Dojo} computers, NVDIA's \textit{Drive Works} platform, or Uber's Michelangelo training framework. However, these are either closed software platforms, built for in-house operations, or are designed to work only on dedicated hardware, as in the case of NVIDIA's software solutions. Based on Elektrobit's history as a key supplier of automotive grade software and real-time operating systems, we have designed EB-AI for interoperability, incorporating optimizations of AI Inference Engines for different embedded ECU devices, coupled with the necessary functional safety standards required by the automotive industry.


	


	
\section{EB-AI Toolchain}
\label{sec:architecture}


In the following, we describe the proposed EB-AI toolchain used to design, train and evaluate the AI Inference Engines. The engines represent application deployment wrappers for trained DNN models, which are used to build driving functions inside EB Robinos$^{\circledR}$, which is Elektrobit's autonomous driving system~\footnote{EB Robinos --- \url{https://www.elektrobit.com/products/automated-driving/eb-robinos/}}.

\subsection{V-Model and Workflow}
\label{sec:v-model}

Automotive software engineering still demands a robust and predictable development cycle. The software development process for the automotive sector is subject to several international standards, namely Automotive SPICE and ISO 26262~\cite{Salay2017}. Accepted standards, as far as the software is concerned, rely conceptually on the traditional V-Model development lifecycle. It is important to approach deep learning from a more controlled V-model perspective in order to address a lengthy list of challenges, such as the requirements for the training, validation and test datasets, the criteria for the data definition and preprocessing, as well as the impact of hyperparameters tuning. 
	
\begin{figure}
	\centering
	\begin{center}
		\includegraphics[scale=0.63]{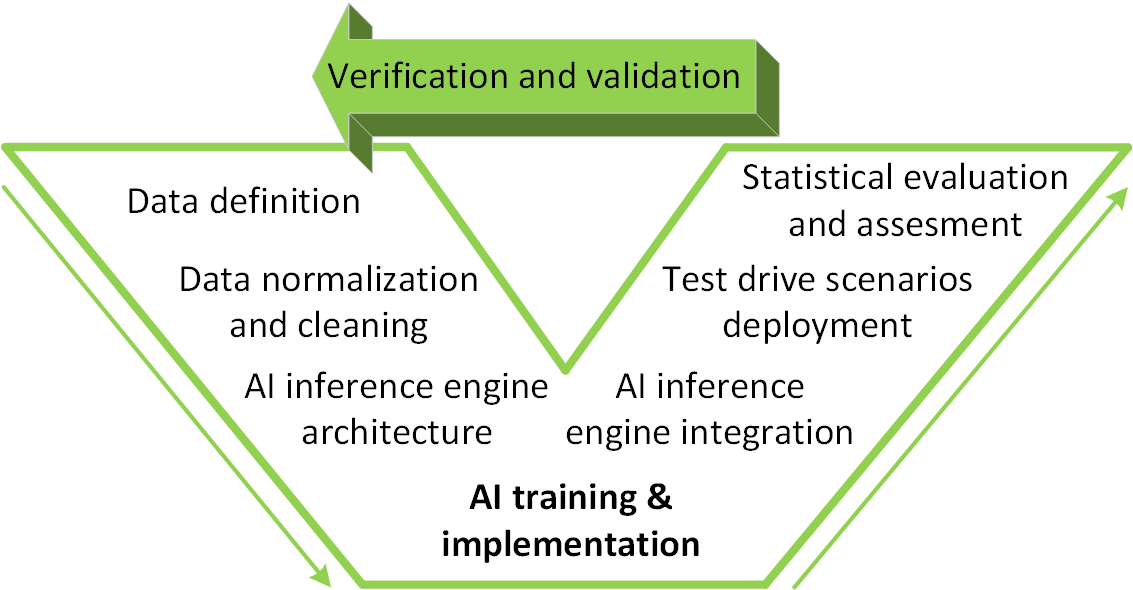}
		\caption{\textbf{Data driven V-Model development of AI Inference Engines}. The proposed V-model encompasses a synergy of data driven SiL and HiL design, training and evaluation of AI components.}
		\label{fig:NN-V-Model-Approach}
	\end{center}
\end{figure}

In Fig.~\ref{fig:NN-V-Model-Approach} we propose a data driven V-Model for prototyping and development within EB-AI. The first two steps regard \textit{data definition} and \textit{data normalization and cleaning}. These are used to define the \textit{AI inference engine architecture}, where we properly configure the layers of the DNN and select the input training data according to the pre-processed datastreams. We then have the \textit{AI training \& implementation} step, where we train the DNN model to the requirements of the application. The trained model is converted automatically to a format that can be deployed within the autonomous vehicle. The converted model represents the \textit{AI Inference Engine} which is integrated and tested in different driving scenarios. According to the results obtained at different levels within the data driven V-Model, we can start the \textit{Verification and Validation} process, where we refine the inference engine.

This circular approach allows the EB-AI architecture to employ continuous learning, where DNNs can improve their accuracy overtime. To perform this task, we proposed a generative technique to compute synthetic data that can be used in the training process based on the GOL algorithm~\cite{grigorescu2018generative}.

	
\begin{figure*}
	\centering
	\begin{center}
		\includegraphics[scale=0.27]{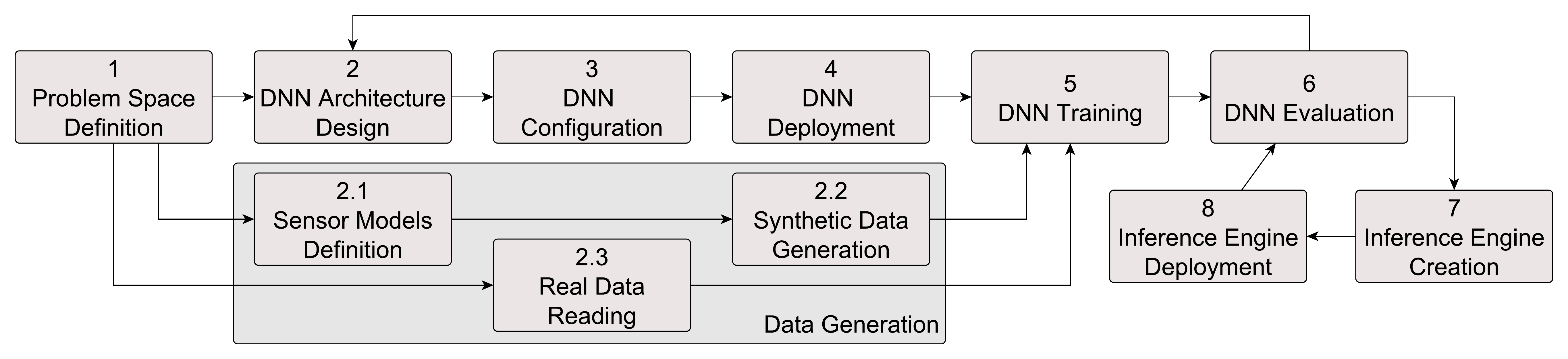}
		\caption{\textbf{AI Inference Engines design, training and evaluation workflow within EB-AI}. The workflow builds on top of the proposed data driven V-model from Fig.~\ref{fig:NN-V-Model-Approach}.}
		\label{fig:eb_ai_flow}
	\end{center}
\end{figure*}
	

Fig.~\ref{fig:eb_ai_flow} details the workflow for AI Inference Engines design, deployment and testing based on the data driven V-Model. Phase 1 consists in defining the problem space and collecting the necessary data for the training process. In this phase, the data is pre-processed, annotated, normalized and filtered. In phase 2 we design the DNN architecture, activation functions, structure of the hidden layers and output nodes. Phase 3 deals with the tuning of the DNN, that is, all hyperparameters necessary for the training phase, such as the learning rate, loss function, regularization strategy and accuracy metrics.

Once the DNN setting has been completed, we deploy the DNNs among the cloud nodes in phase 4 and start the training in phase 5. In this phase we feed the DNN both with real sensory data, as well as with synthetic data, generated in a parallel workflow branch (phases 2.1-2.3). After training, the DNN is evaluated and redesigned if necessary (phase 6). The output of the training and evaluation phases is the \textit{AI Inference Engine} (phase 7). This is then deployed among the edge device in the car (phase 8) and evaluated again with respect to in-car performance measures.
	

\subsection{EB-AI Architecture}
\label{sec:detailed_architecture}
	
	
\begin{figure}
	\centering
	\begin{center}
		\includegraphics[scale=0.3]{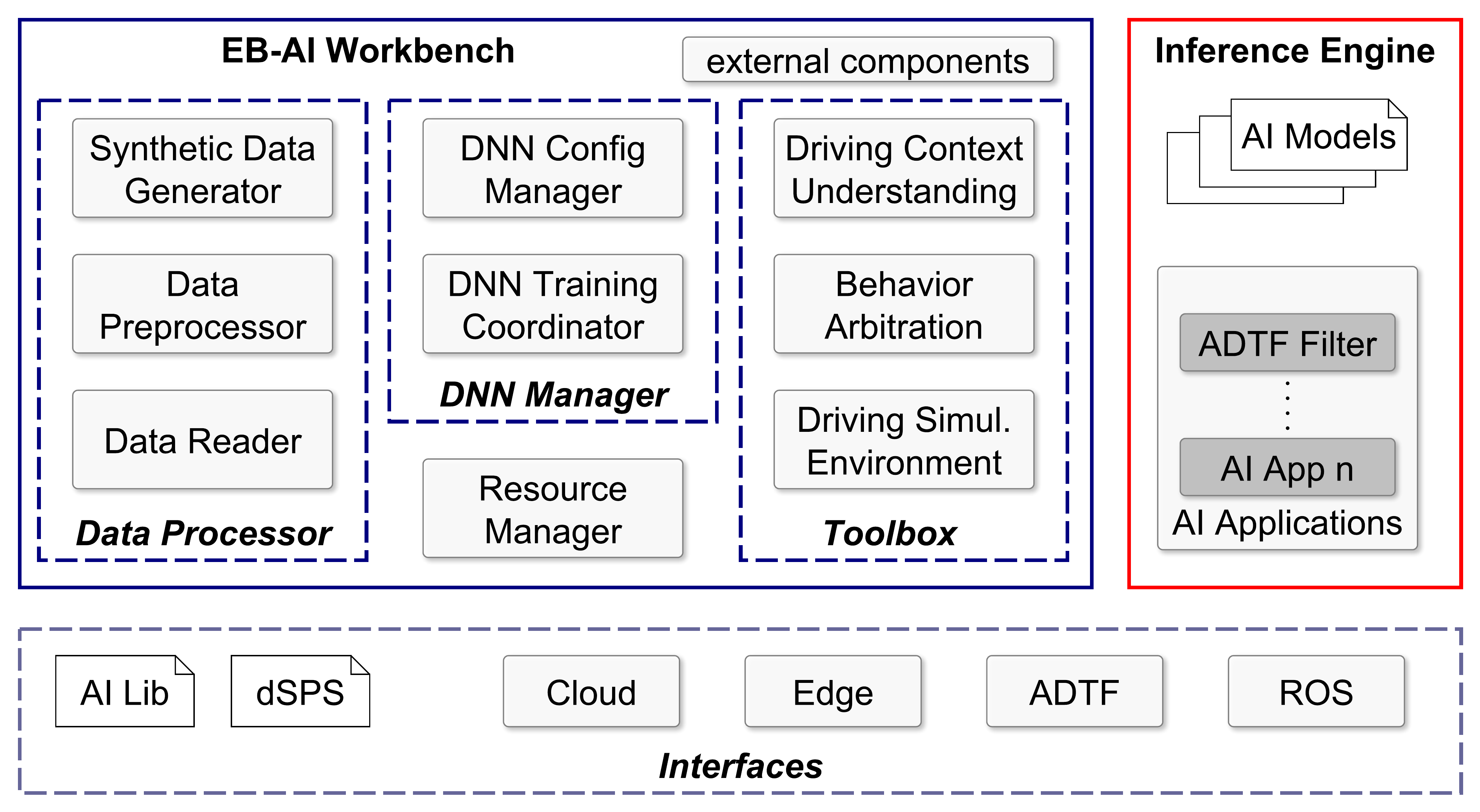}
		\caption{\textbf{EB-AI cloud architecture}. The framework follows the two major operations from the DL workflow: design and training in the EB-AI Workbench and inference. The Interfaces allow for flexibility with respect to the core AI libraries, edge device hardware and type of application deployment wrapper.}
		\label{fig:eb_ai_simplified_architecture}
	\end{center}
\end{figure}

The EB-AI cloud architecture is depicted in Fig.~\ref{fig:eb_ai_simplified_architecture}. It is based on two main components: \textit{i}) the \textit{EB-AI Workbench} and \textit{ii}) the \textit{Inference Engines}. The former generates and trains AI models, while the latter uses such models for prediction and inference on different embedded ECUs.

Specifically, the EB-AI Workbench is responsible for prototyping, training, and development of the AI models and is composed of four main modules: \textit{i}) \textit{Data Processor}, \textit{ii}) \textit{DNN Manager}, \textit{iii}) \textit{Resource Manager} and \textit{iv}) \textit{Toolbox}.
	
\smallskip
\noindent{\textbf{Data Processor.}} It reads, pre-processes and analyzes the data provided by the vehicles' sensors. Data pre-processing typically deals with detecting outliers in the input data, visualizing it using feature space reduction techniques such as PCA (Principal Components Analysis) and t-SNE (t-distributed Stochastic Neighbor Embedding), checking its variance and standard deviation across its data distribution domain, splitting it into appropriate train-evaluation-test batches and finally normalizing it. The probable capacity of the DNN model is determined based on the quantity of training samples. Furthermore, this stage is used to generate synthetic data through our patented GOL algorithm~\cite{grigorescu2018generative}. This allows to learn patterns of real data and generate artificial samples to improve the data available when training the DNN.

\smallskip
\noindent{\textbf{DNN Manager.}}
It contains two modules: the \textit{DNN Config Manager} handles the configuration of the DNN, while the \textit{DNN Training Coordinator} (TC) manages the training process. Similar to the work in~\cite{Villalonga_2020}, DNN models are automatically configured based on the results obtained from pre-processing the training data. By analyzing the structure and type of input data, the DNN Config Manager automatically determines the input and output shapes of the DNN architecture, thus enabling the usage of existing network architectures from the workbench. 
The first step is to automatically define the input and output layers of the model based on the input-output data types. The input layer's shape is given by the structure of the training samples (e.g. images, LiDAR, radar, ultrasonic and IMU measurements, log traces, GPS tracks, etc.), while the output layer's shape is computed based on the labels' type. For example, if the training data is composed of image samples and 2D bounding boxes, then different network architecture for spatial object recognition are suggested, as shown in use-case~\ref{sec:use_cases_env_percep}. Similarly, as presented in subsection~\ref{sec:mpp}, different RNN architectures are proposed if the data is structured as temporal sequences. In this first step, training samples and labels are aggregated into single input and output layers, respectively. The automatic configuration of the input-output layers can be manually tuned if the samples and labels are composed of multiple types of sensory measurements (e.g. images and LiDAR) and/or multiple prediction heads (e.g. 2D object detection and 3D object reconstruction).

\noindent Once the input-output shapes of the DNN model have been defined, we proceed to computing the architecture of the inner layers. We split the anatomy of a DNN architecture into three main blocks: \textit{i}) backbone model, \textit{ii}) feature extractor model and \textit{iii}) prediction heads. These are chosen from subsets of state-of-the-art DNN architectures, such as VGG16, MobileNetV2 and Darknet for the backbone, or YoloV3-V5, RetinaNet and SSD (Single Shot Detector) for the case of prediction heads in object detection. We compute the DNN structure depending on the model capacity determined at Data Processor stage. Namely, in order to avoid overfitting if training data is scarce, we use a light-weighted backbone, such as VGG16, coupled to prediction heads having a lower number of convolutional layers. Different DNN models are suggested based on this analysis. The models are subsequently adapted and modified in our SiL and HiL paradigm for obtaining an optimal AI Inference Engine for the task at hand.

\noindent The TC employs libraries for DL on top of a Distributed Stream Processing System (dSPS), allowing us to run a distributed \textit{training application} on the dSPS. {\newcolor{blue}Such an application} is represented as a directed acyclic graph, where vertices represent \textit{operators} and edges represent streams between pairs of operators. Each operator carries out a piece of the overall computation, that is a \textit{training task}, and can be distributed over multiple instances.


\begin{figure}
	\centering
	\begin{center}
		\includegraphics[scale=0.5]{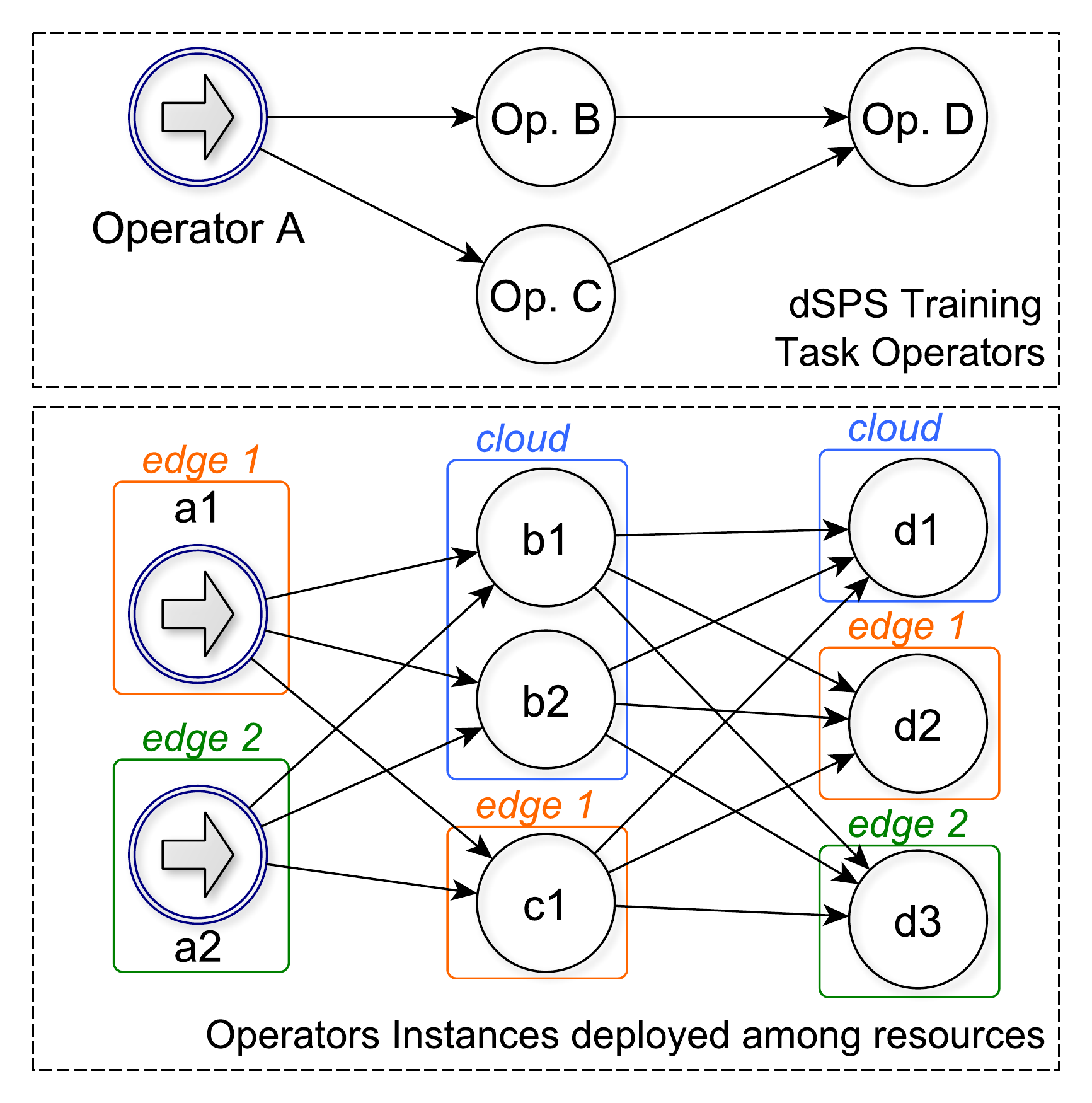}
		\caption{\textbf{Training Tasks Operators of the dSPS}. In this example we show an instance where data managed by the operator \textit{A} is sensitive, thus the RM deploys operator instances \textit{a1} and \textit{a2} only on the two edge nodes. Operator \textit{B} can instead be performed in the Cloud, while Operator \textit{C}, which produces the AI Inference Engine, is available on all nodes.}
		\label{fig:training-task-dsps}
	\end{center}
\end{figure}

\smallskip
\noindent{\textbf{Resource Manager.}} 
It {\newcolor{blue}is the elastic manager that dynamically} handles the deployment of the DNN training tasks over cloud and edge resources.
Fig.~\ref{fig:training-task-dsps} shows an example of training tasks for a dSPS application composed of four operators (A, B, C, D) allocated over the cloud and two edge nodes.

The Resource Manager (RM) is able to reconfigure over time \textit{i}) the number of edge/cloud resources, \textit{ii}) the operators' parallelism and \textit{iii}) the allocation of operator instances to available resources. Thus, RM can redistribute dynamically the training tasks on a different pool of resources to keep desired performances according to the input workload (e.g. the input rate of values read by the vehicles). The operators of the dSPS application can be distributed among edge and cloud nodes according to how {\newcolor{blue}much} resources are necessary for tasks, as well as according to the sensitivity of the acquired training data, ensuring that sensitive data is processed where it is produced, without propagating it to other cloud or edge nodes, thus mitigating privacy. 
Once the learning is completed, the trained model is replicated among all edge and cloud nodes so that vehicles can immediately have the latest version available. 
	
\smallskip
\noindent{\textbf{Toolbox.}} 
It is a set of components specific for automotive AI applications. The training itself feeds the previously defined data to the network, allowing it to learn a new capability by reinforcing correct predictions and correcting the wrong ones. The module \textit{Driving Context Understanding} classifies the driving context from grid-fusion information, the \textit{Behavior Arbitration} understands the driving context and optimizes the strategy from real-world grid representations, while the \textit{Driving Simulation Environment} trains, evaluates and tests AI algorithms in a virtual simulator, such as Microsoft's AirSim.

\smallskip
\noindent{\textbf{Inference Engine.}} 
It contains the trained AI models from the EB-AI Workbench and deployment wrappers (or application) for the models. For example, such a wrapper can be built as a Robotic Operating System (ROS) Node~\footnote{ROS --- \url{https://www.ros.org/}}, or an EB Assist ADTF$^{\circledR}$ component. ADTF stands for \textit{Automotive Data and Time-Triggered Framework}~\footnote{ADTF --- \url{https://www.elektrobit.com/products/automated-driving/eb-assist/adtf/}} and is a tool for the development, validation, visualization and testing of advanced driver assistance systems and automated driving features.

\smallskip
\noindent{\textbf{Interfaces.}} 
Both EB-AI Workbench and Inference Engine use an interface layer for low-level libraries (e.g. CUDA, Tensorflow, PyThorch, CNTK, or Caffe2), integration with the dSPS engine (e.g. Spark) and the APIs for cloud and edge nodes. Furthermore, it provides a specific interface to ADTF and ROS.


\section{Hybrid Deployment Advantages}
\label{sec:cloud_edge}
	
This section discusses the advantages brought by the proposed hybrid deployment. Specifically, it discusses how our solution preserves privacy (Section~\ref{sec:privacy}) and how it improves fault tolerance and scalability (Section~\ref{sec:hybrid}). In Appendix \ref{appendixA} we also provide an experimental evaluation aimed to show the benefit of increases parallelism in DNN training.

\subsection{Privacy-preserving Techniques}
\label{sec:privacy}
	

Although the Cloud can offer a scalable training infrastructure, some training data should not flow to the Cloud due to data protection requirements, as those on personal data enforced by the GDPR. Both data encryption and obfuscation solutions can be used to address these challenges.

Data obfuscation techniques like Differential Privacy (DP) have been successfully used both in centralized and federated learning~\cite{3324405}. Differently from pre-processing techniques, using DP on Edge devices as part of a federated learning approach guarantees better accuracy~\cite{HuangMFLG17,YangLCT19}. In particular, Edge devices are carrying out the training of a number of front layers by using DP on the sensitive data. Then, they transmit the intermediate results to the Cloud for the completion of the training and dissemination to the Edge of the updated layers.

Learning on encrypted data has also been proposed. While homomorphic encryption has been prototyped on centralized Cloud training~\cite{6410315}, applications of Multi Party Computation (MPC) offers distributed privacy-preserving training~\cite{7958569}.

The EB-AI toolchain supports DP learning by relying on the TensorFlow Privacy library\footnote{TensorFlow Privacy --- \url{https://github.com/tensorflow/privacy}}. Differently from MPC learning that must rely on ad-hoc computing framework (e.g. Sharemind~\cite{Bogdanov:2008}), DP learning can be managed by the EB-AI framework to take advantage of Edge devices on self-driving cars. 
The TensorFlow Privacy library relies on a differentially private stochastic gradient descent (SGD) algorithm~\cite{abadi2016deep}, which provides privacy protection for deep neural networks while incurring a tolerable overhead in terms of training time.


	

{\newcolor{blue}		
\subsection{Fault Tolerance and Elastic Scalability}
\label{sec:hybrid}
}
The EB-AI architecture can support flexible learning deployment across Cloud and Edge, offering scalable execution of learning tasks according to resource and privacy constraints. 

Despite the advances in OTA software, the deployment of learning tasks must tolerate faults due to lack of connectivity with AVs. 
Distributed computing frameworks, such as Apache Spark, can be deployed to enable {\newcolor{blue}elastic} fault tolerant federated learning via, for example, TensorSpark and CaffeOnSpark. 

{\newcolor{blue}
The EB-AI Resource Manager extends the ELYSIUM~\cite{lombardi2017elastic} autoscaler in order to 
}
rebalance the operators of the training application. {\newcolor{blue} Specifically, it elastically scales both the dSPS operators' parallelism and resources (cloud and edge nodes), allocating these operators according to workload changes or faults predicted and/or observed.}

The flexibility of the task definition provided by existing dSPS platforms, together with their seamless integration with high-level AI frameworks, allows the EB-AI workbench to directly introduce new pre-processing or pre-training tasks at the Edge according to emerging privacy and learning constraints. 

	
	
	
	

\section{EB-AI Use Cases}
\label{sec:use_cases}

State of the art autonomous driving systems are defined using modular perception-planning-action pipelines, in which the main problem is divided into smaller sub-problems, each module being designed to solve a specific task and deliver the outcome as input to the adjoining component~\cite{Grigorescu_JFR_2020}.

In the following, we aim to highlight the functionalities of the proposed EB-AI framework in two autonomous driving application use-cases, namely \emph{driving environment perception} and \emph{most probable path prediction}. Since this article focus on the underlying computing framework, as opposed to a specific algorithm, we will not focus on the accuracy of the two use-case methods, but on their designed, deployment and evaluation using the proposed EB-AI framework.

In both use cases presented below, we use the proposed Elastic Cloud2Edge paradigm to perform data ingestion and the configuration of the DNNs using EB-AI's Workbench interface from Fig.~\ref{fig:EB_AI_pipeline}, while the \textit{DNN Manager} and \textit{Resource Manager}, detailed in Section~\ref{sec:architecture}, are used to jointly distribute the training on the Cloud and the AVs' Edge devices.

\subsection{Use Case 1: Driving Environment Perception}
\label{sec:use_cases_env_percep}

The perception of the driving environment is a key part of any autonomous driving system, characterized as the ability of the inference engine to understand the image scene and accurately represents the dynamic and stationary objects (e.g. cars, pedestrians, traffic signs, etc.). The EB-AI framework offers ready to use DNN architectures, such as SegNet for traffic scene image segmentation~\cite{badrinarayanan2015segnet}, different versions of the Yolo detector~\cite{RedmonYOLO2016} and our own Deep Grid Net (DGN) for driving context understanding~\cite{marina2019dgn}. For the use case of environment perception, the specific problem space of object detection in images will be considered.

\begin{figure}
	\centering
	\begin{center}
			\includegraphics[scale=1.2]{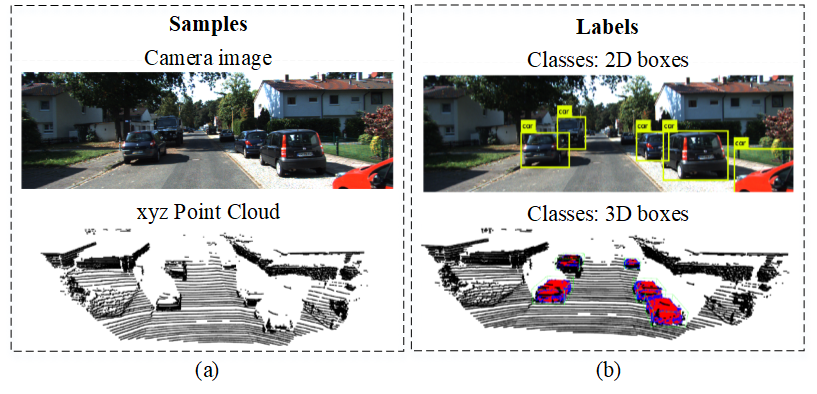}
			\caption{\textbf{Pairs of samples and labels commonly used in environment perception DNNs.} a) Input RGB image and 3D point cloud. b) Labels of objects of interest annotated in 2D on images and in 3D onto point cloud data, respectively.}
			\label{fig:env_perception}
	\end{center}
\end{figure}

For environment perception, we perform object detection and classification on RGB color images and XYZ point clouds, as illustrated in Fig.~\ref{fig:env_perception}. We used labelled data from the CamVid dataset \cite{CamVid2018}, which we have converted into the ROS bags. The ROS bags are uploaded into the EB-AI Cloud for pre-processing operations such as normalization, encoding or clipping.

\begin{figure}
	\centering
	\begin{center}
		\includegraphics[scale=0.92]{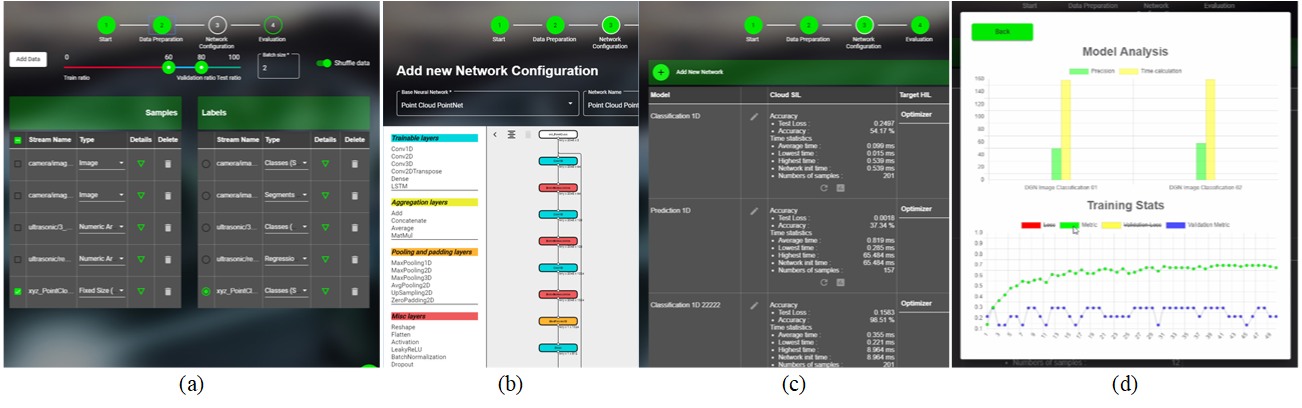}
		\vspace{0.5em}
		\caption{\textbf{Snapshots from the EB-AI Cloud interface.} (a) Data upload and pre-processing. (b) DNN architecture design. (c) Deployment and evaluation in the Cloud and Edge ECU as SiL and HiL, respectively. (d) Models comparison.}
		\label{fig:EB_AI_pipeline}
	\end{center}
\end{figure}

Fig.~\ref{fig:EB_AI_pipeline} shows a couple of snapshots from the EB-AI Cloud interface, depicting the data upload, DNN architecture design and training as SiL, deployment and evaluation via HiL and models comparison. The first steps in building an AI inference engine is the definition of the requirements and collection of the necessary data for training. The data itself is collected from two sources, namely vehicles' sensors and synthetic generation through our GOL algorithm~\cite{grigorescu2018generative}. This allows the improvement of the patterns learned from real data and the ability of the trained network to generalize. A manual splitting ratio of the data between train, validation and test is available, as well as automatic k-fold variations.

By analyzing the input data the \textit{DNN Config Manager} automatically determines the input and output shape of the DNN architecture, thus enabling the usage of existing network architectures from the workbench. Additional adjustments can be made in order to improve the accuracy of the DNN, based on the task at hand. In this driving environment perception use-case we use a Yolo V5 DNN trained on pairs of camera images and annotated 2D bounding boxes of objects of interest.

The training is managed by the \textit{DNN Training Coordinator}, which handles the DL libraries on top of a distributed streaming processing system. The Resource Manager handles the available training resources over the cloud and/or edge device. In order to illustrate the fault tolerance and scalability of the training process, we have used different ECU configurations with 8, 12 and 16 nodes respectively. The ECUs are a baseline desktop computer equipped with an Intel Core i9 9900K CPU, 64 GB RAM, with two high-performance NVIDIA GeForce RTX 2080 Ti graphics cards, an NVIDIA Jetson AGX platform, a Kalray Konic board and a RaspberryPi 3B+ computer.

\begin{figure}
	\centering
	\begin{center}
		\includegraphics[scale=0.42]{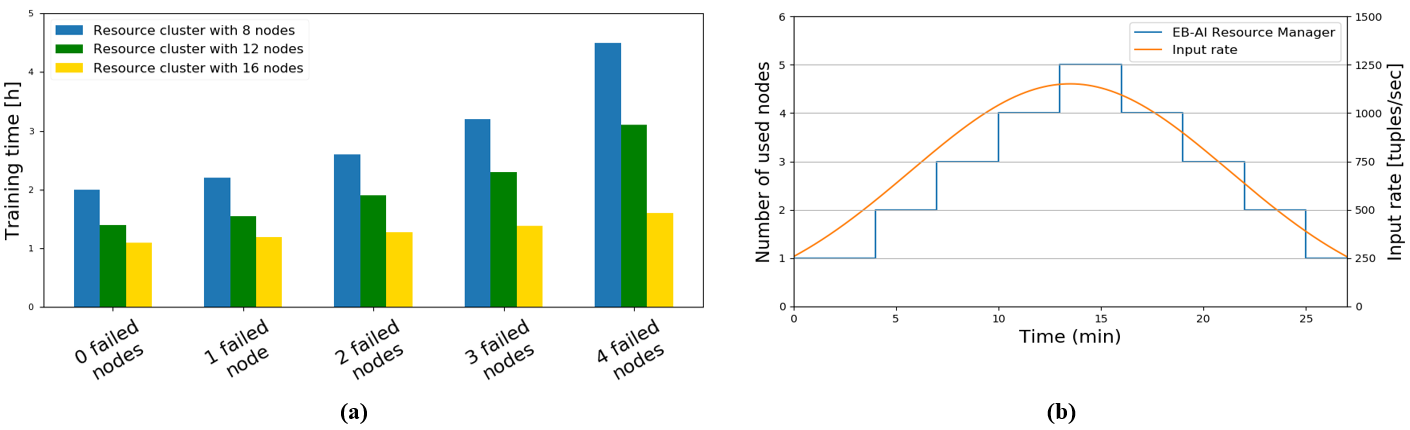}
		\caption{\textbf{Usage of training resources in the context of fault tolerance and elastic scalability.} (a) Fault tolerance based on number of failed nodes. (b) Elastic scalability of the EB-AI Resource Manager.}
		\label{fig:fault_tolerance_scalability}
	\end{center}
\end{figure}

Fig.~\ref{fig:fault_tolerance_scalability} shows how the training resources are used in the context of fault tolerance and elastic scalability. In order to evaluate the fault tolerance, specific resource nodes were deliberately blocked, while the impact on the overall training time was measured. As it can be seen in Fig.~\ref{fig:fault_tolerance_scalability}(a), with a hardware setup comprised of 16 nodes, the impact of failing 1 to 4 nodes is negligible on the training time. However, the impact becomes visible when 4 nodes (50\% of total available nodes) are blocked in a hardware setup of 8 total nodes. Fig.~\ref{fig:fault_tolerance_scalability}(b) shows the number of used nodes in the training process while injecting an input stream curve of 250-1200 tuples (subset data) per second. The EB-AI Resource Manager reactively allocates or deallocates nodes in order to compensate the input demand.

\begin{figure}
	\centering
	\begin{center}
		\includegraphics[scale=0.25]{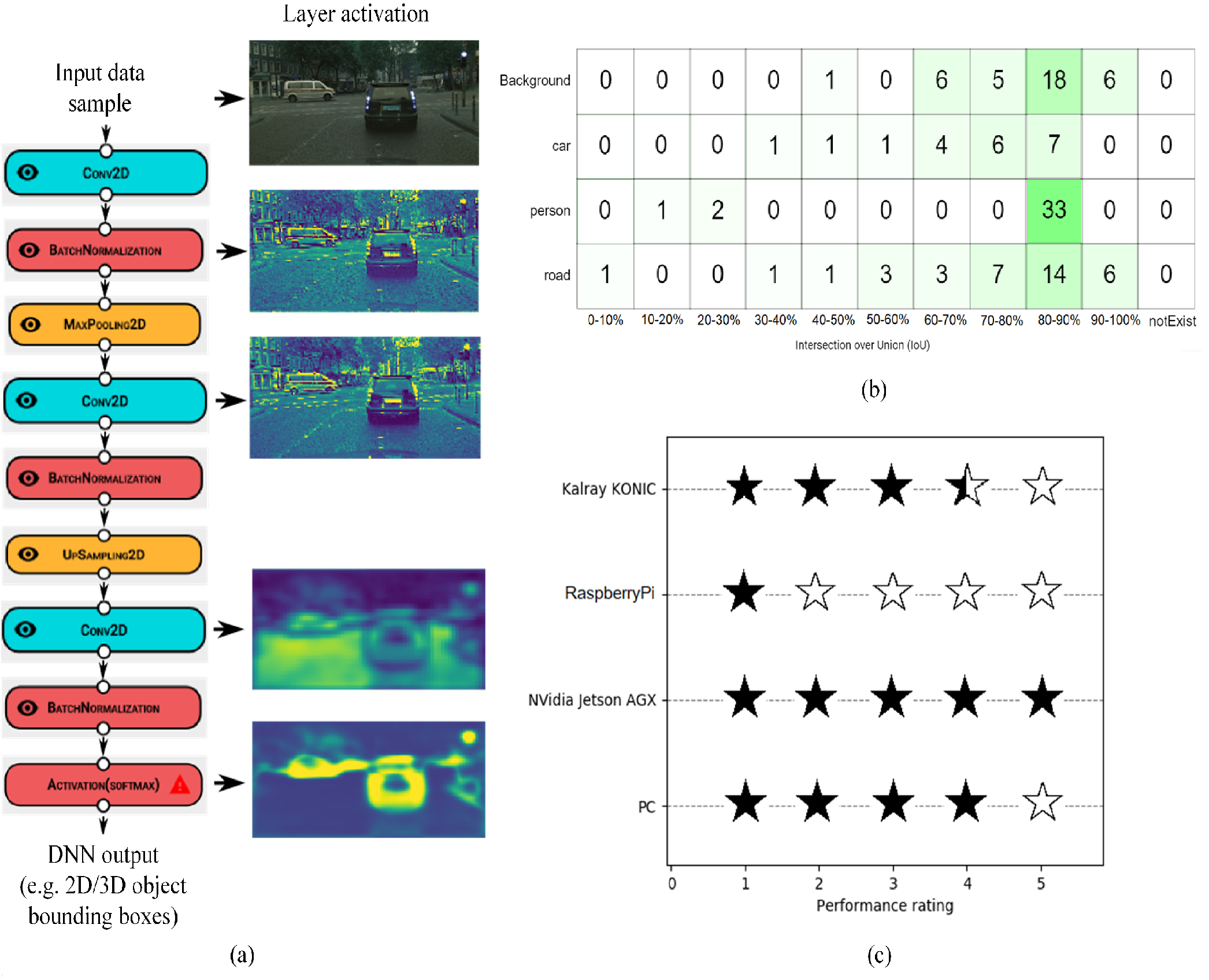}
		\caption{\textbf{Performance evaluation metrics within EB-AI.} (a) Direct visualization of the activations of each DNN layer. (b) Intersection over Union (IoU) evaluation matrix describing in percentage the amount of overlapping bounding boxes. (c) 5-starts rating aggregates overall accuracy.}
		\label{fig:metrics_perception}
	\end{center}
\end{figure}

The performance of a DNN can be evaluated based on a series of metrics, depending on its architecture. Fig.~\ref{fig:metrics_perception} show different evaluation approaches for our object detection use-case based on direct visualization of the DNN layers' activations, Intersection over Union (IoU), as well as a performance estimator which aggregates the overall accuracy of the model into a 5-stars rating system.

To assess the precision of object detection, we use the IoU quality measure, which is a percentage measure reflecting the amount of overlapping between the predicted bounding box, as outputted from the AI inference engine, and the labeled bounding box. An accurate model has an IoU value close to 100\%, which represents maximum accuracy. Fig.~\ref{fig:metrics_perception}(b) illustrates the IoU matrix determined for the trained Yolo V5 model. From the obtained IoU matrix, it can be observed that the majority of the detected objects overlap in a ratio of 80-90\% with the ground truth. Featuring easy interpretability over the testing data, the user can trace back the samples where the evaluation failed or was not accurate enough, while iteratively improving the neural network until the desired performance is achieved.

\begin{figure}
	\centering
	\begin{center}
		\includegraphics[scale= 0.48]{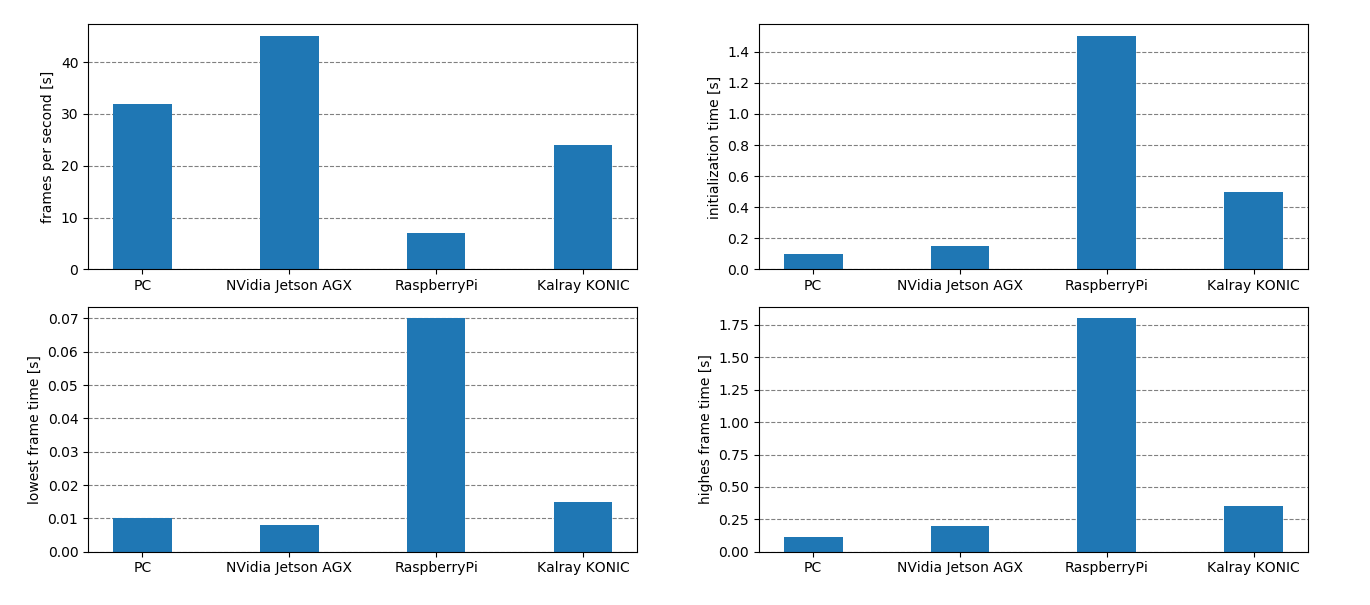}
		\caption{\textbf{Real-time performance metrics obtained on different deployment ECU devices}. The diagrams show the Frames per Second (FPS), AI Inference Engine initialization time, as well as the lowest and highest achieved framerate for four different ECUs.}
		\label{fig:eb_ai_obj_det_perf_statistics}
	\end{center}
\end{figure}

The obtained AI Inference Engine contains the trained DNN model encapsulated in a deployment wrapper. In this case, it was deployed as an ADTF wrapper on Elektrobit's Automotive Data and Time Triggered Framework (ADTF). Fig.~\ref{fig:eb_ai_obj_det_perf_statistics} shows different measure performance indicators for the four considered target ECU devices. The metrics quantify the computation time in frames per second, initialization time, as well as maximum and minimum computation time, respectively. The obtained results are proportional to the computation power of each ECU, yielding low framerates for low-range grade devices, such as the RaspberryPi, as opposed to the high performance of automotive ECUs, such as the NVIDIA AGX platform. The metrics are logged in the ECU device for latter importing into the Cloud, where they optimized by reconfiguring and retraining the DNN using the proposed data driven V-Model.

\subsection{Use Case 2: Most Probable Path Prediction}
\label{sec:mpp}

\begin{figure}
	\centering
	\begin{center}
		\includegraphics[scale=0.455]{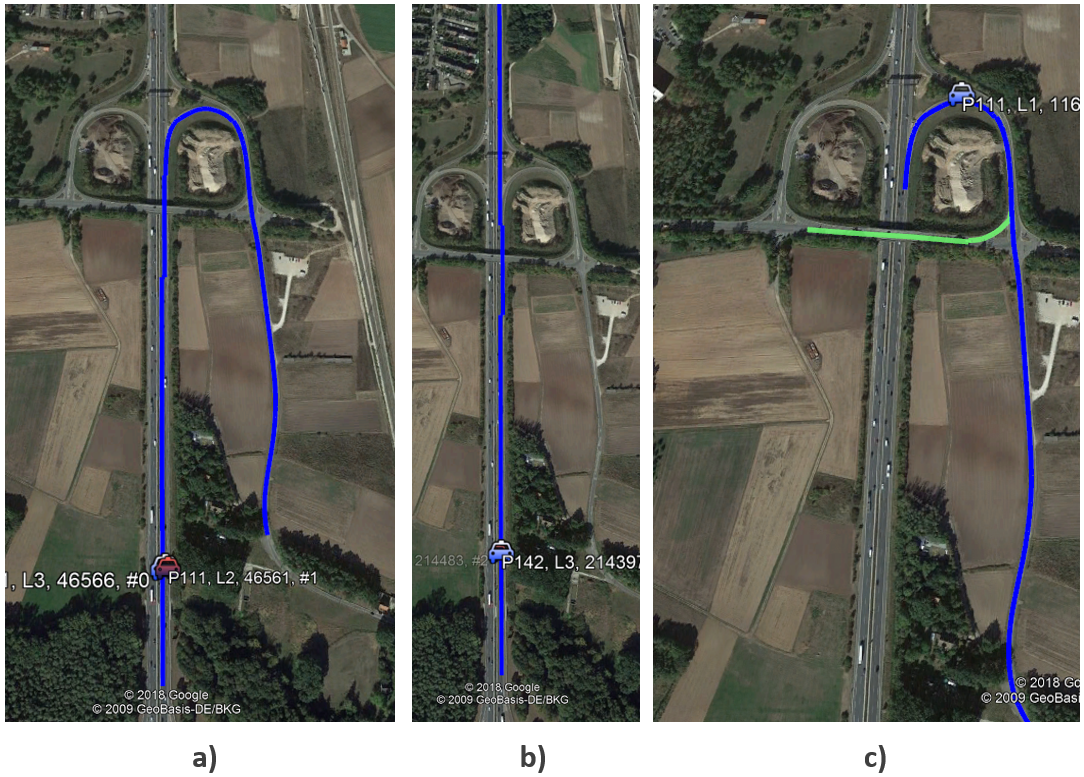}
		\caption{\textbf{Comparison between different MPP computation methods}. a) Ground truth. b) Prediction given by a classical (non-AI) method. c) Correctly predicted path using an AI Inference Engine.}
		\label{fig:mpp_results}
	\end{center}
\end{figure}

In autonomous driving and driver assistance, in order to prepare the vehicle for upcoming hazards or to warn the driver, it is important to predict in advance the route that the vehicle will follow over a future time horizon. Such algorithms, illustrated in Fig.~\ref{fig:mpp_results}, are called Most Probable Path (MPP) estimators. With an increasing level of map details, past trips of vehicles can be used to store information about the road paths that are likely to be reached within a specified distance using GPS observations. Using this information, the MPP algorithm can learn and further predict the probable road network ahead. As the vehicle travels forward, travelled road segments are left behind, while new ones are added to the MPP.

A map is a web of paths implemented using a tree structure, where each node in the tree represents a link between two segments of the road. The root node of the tree represents the link where the vehicle is currently located. Starting from a link, the vehicle can travel to a considerably large number of possible paths. A transition from one link to a reachable successor is represented by adding a child to the proper parent node in the tree. This is important because as the vehicle travels forward, nodes that become unreachable are removed from the tree. Every time the vehicle transitions from one link to the next, at least one node becomes unreachable and will be removed from the graph. 

In order to predict where the vehicle is heading in the future, the MPP method needs to calculate the likelihood associated to each possible turn. In the classical way, the weight associated to each turn, or transition, depends on many factors and considers the turning angle and the road class of the following road segment. A smaller turning angle results in a higher probability of driving straight, whereas a bigger turning angle results in a smaller probability of following that specific road segment. 

The main drawback of the classical approach in computing the MPP is that in most cases these predictions are performed at a small distance (a few hundred meters) in front of the current position, since the computation time required for computing the MPP probabilities is high. Additionally, the result will always be similar, since the turning angle and road class will be constant on the same map, so there will never be any difference between different drivers or different cars.

\begin{figure}
		\centering
		\begin{center}
			\includegraphics[scale=0.5]{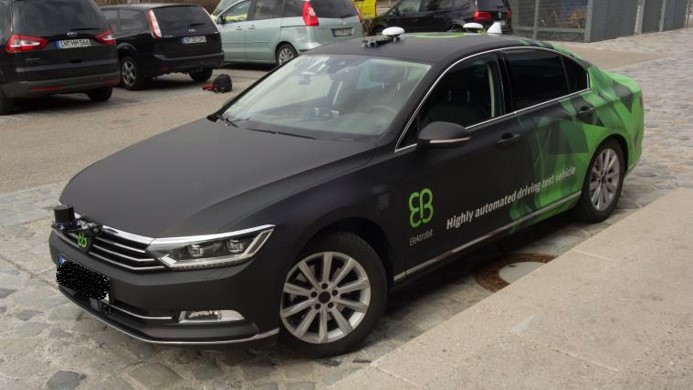}
			\caption{\textbf{Test vehicle used for data acquisition and testing for the MPP AI Inference Engine.}}
			\label{fig:mpp_car}
		\end{center}
\end{figure}

In this use-case, we have trained and deployed an AI Inference Engine using GPS data collected with Elektrobit's autonomous test vehicle, shown in Fig.~\ref{fig:mpp_car}. We have performed $25$ trips of $40km$ each. Using the EB-AI framework, we were able to enhance the links using the pre-processing operations. As in the environment perception use-case example, the user can split the ratio of training data between train, validation and test datasets.

The DNN at the core of the AI Inference Engine is a Recurrent Neural Network (RNN) consisting of $256$ Long-Short-Term-Memory (LSTM) units, followed by a Fully Connected layer (SoftMax activation), providing as output a list of predictions of the most probable paths. We have chosen an RNN network architecture due to its robustness in processing time dependent sequences, thus encoding the links relations in its memory cells. EB-AI was used to efficiently design and optimize the neural network. The obtained AI model wrapped in the ADTF format was deployed in a container ready to run on a NVidia Jetson AGX board mounted in the vehicle from Fig.~\ref{fig:mpp_car}.

\begin{figure}
		\centering
		\begin{center}
			\includegraphics[scale=0.62]{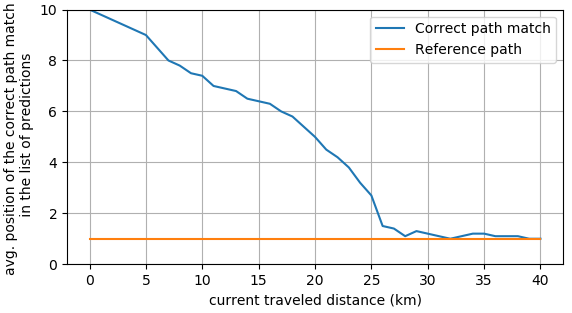}
			\caption{\textbf{The average correct route match relative to the travelled distance.} At halfway through the $40km$ trip, the predicted route is within the top 5 matches, while in the last $10km$ the predicted route is the top match.}
			\label{fig:mpp_perf_rez_avg_corr_route}
		\end{center}
\end{figure}

To evaluate the performance of the MPP approach, EB-AI provides the means to measure the similarity between two paths, namely predicted and ground truth. The output of the DNN is a list of paths ordered by the probability of being the correct one. Fig.~\ref{fig:mpp_perf_rez_avg_corr_route} shows the average position of the most probable route as the trip advances. The $y$ axis in the figure represents the ground-truth reference path and the DNN predicted path, order based on their probabilities. We show the top most probable paths estimated by the neural network. Ideally, the predicted path should be the same as the reference one. In the beginning, when little information is known about the past vehicle's route, the MPP has a lower accuracy. The accuracy increases gradually as the drive progresses, reaching top match after kilometer $25$.

\begin{figure}
	\centering	
	\subfloat[]{\label{fig:mpp_corr_pred_route}\includegraphics[scale=0.5]{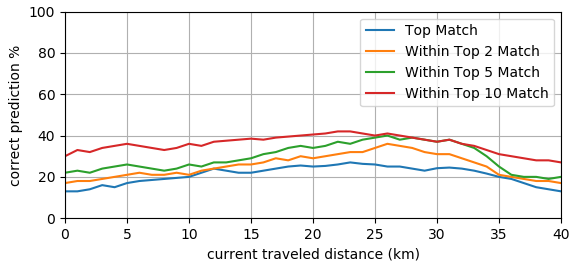}}
	\subfloat[]{\label{fig:mpp_corr_pred_repeat}\includegraphics[scale=0.515]{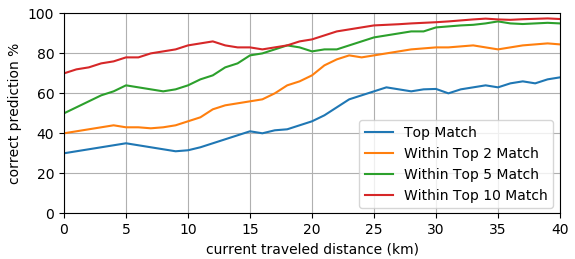}}
	\caption{\textbf{Performance of the MPP algorithm across different trips.} (a) Driving on new routes. (b) Driving repetitively on the same routes.}
	\label{fig:mpp_perf_results}
\end{figure}

An additional performance metric is the correct prediction rate of the DNN. Fig.~\ref{fig:mpp_perf_results} presents how often the MPP correctly predicted across different trips. The left most graph presents the results obtained when driving all the time on new routes, while the right most graph illustrates the results obtained when driving repetitively on the same routes.

\section{Conclusions}
\label{sec:conclusions}

In this paper, we have proposed an elastic AI development framework for autonomous driving applications based on deep learning, which takes into account some of the advances brought about by AI, Cloud and Edge computing. A modular Elastic toolchain providing all the required DL components enables the deployment of training tasks over both Cloud end Edge resources, where the latter can be located on the vehicles themselves. Pre-processing raw data and/or training partial models at the Edge allows to reduce the amount of data to upload to the Cloud and help mitigate privacy issues. We showed the effectiveness of the proposed toolchain in two application cases implemented at Elektrobit Automotive, namely Environment Perception and Most Probable Path Prediction. Furthermore, we showed the convenience of exploiting a distributed setting to parallelize as much as possible the training of the DNNs.

As future work, we aim to prototype and deploy AI Inference Engines at a large scale using the proposed hybrid deployment strategy over a larger fleet of autonomous vehicles.

\appendixtitles{no} 
\appendix
\section{}\label{appendixA}
\unskip


A hybrid deployment over cloud and edge resources allows to significantly increase the level of parallelism in AI Inference Engine development.
To give an idea about the scale of improvement that parallelisation can bring to DNN training, we present a toy example with
an experimental evaluation where two DNNs, i.e., a Multi-Layer Perceptron (MLP) and a CNN, are trained. The MLP has 2 hidden layers of 800 units each, followed by a softmax output layer of 10 units with 20\% dropout to input data and 50\% dropout to both hidden layers. The CNN has 2 convolutional layers with 32 kernels of size 5x5 strided and padded, a fully-connected layer of 256 units with 50\% dropout on its inputs and a 10-unit output layer with 50\% dropout on its inputs.
We compare the time required to train them by using a CPU, i.e. low level of parallelism, and a GPU, higher parallelism.



The experimental evaluation was carried out on a Dell Precision T7610 server, equipped with a Intel Xeon E5-2660 v2 CPU, 8GB of RAM, and with two GPUs, namely Tesla K20c and Nvidia Quadro K4000. The DNNs were developed in Python with Lasagne, by relying on Theano to exploit the CUDA GPU libraries. 
The used dataset is based on MNIST, with a training set of 60k samples and a test set of 10k samples. 
%
The tests aim at comparing: i) the training accuracy of the MLP and the CNN, after the same number of training iterations and ii) the training time on the available CPU and GPUs. Optimal settings for each network were computed based on a Q-Learning approach~\cite{lombardi2017elastic} that can automatically apply the dropout to minimize overfitting.





\begin{figure}
	\centering
	\begin{center}
		\includegraphics[scale=0.60]{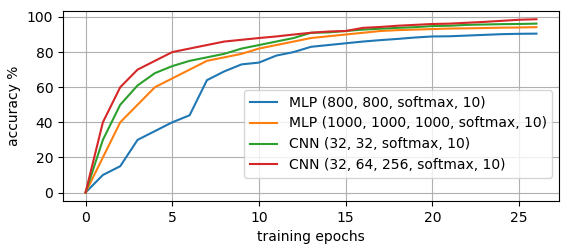}
		\caption{\textbf{Accuracy comparison between MLP and CNN.}}
		\label{fig:accuracy}
	\end{center}
	\vspace{-1em}
\end{figure}

\begin{figure}[t]
	\centering	
	\subfloat[Comparison for training an epoch of MLP.]{\label{fig:time-mlp}\includegraphics[scale=0.65]{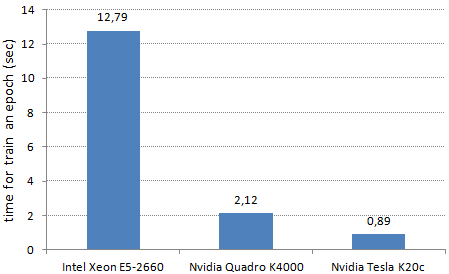}}
	\subfloat[Comparison for training an epoch of CNN.]{\label{fig:time-cnn}\includegraphics[scale=0.65]{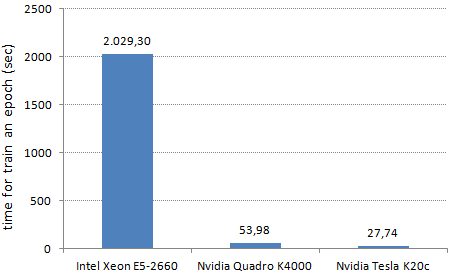}}
	\caption{\textbf{Performance evaluation of the two types of DNNs}}
	\label{fig:reg_GFS_GFPNet}
\end{figure}

Fig.~\ref{fig:accuracy} reports the accuracy results for the two networks during training. The CNN achieves better accuracy than the MLP, being over 98\% in 5 epochs, while the MLP requires about 50 epochs. The maximum accuracy of the CNN is 99.21\% versus 98.56\% of the MLP.
Fig.~\ref{fig:time-mlp} and \ref{fig:time-cnn} report, respectively for MLP and CNN, the performance of training an epoch on the different hardware devices. As expected, training with both the GPUs is much faster than with the CPU. The epoch training time for MLP (resp., CNN) on the GPU Tesla and Quadro are $\approx 14$ and $\approx 6$  (resp., $\approx 73$ and $\approx 38$) times faster than on the CPU. 
Notice how the Tesla is about 2 times faster than the Quadro. 





\authorcontributions{Conceptualization, Sorin~Grigorescu, Bogdan~Trasnea and Federico~Lombardi; methodology, Sorin~Grigorescu, Tiberiu~Cocias, Bogdan~Trasnea, Andrea~Margheri, Federico~Lombardi and Leonardo~Aniello; software, Sorin~Grigorescu, Tiberiu~Cocias and Bogdan~Trasnea; validation, Sorin~Grigorescu, Tiberiu~Cocias, Bogdan~Trasnea and Federico~Lombardi; formal analysis, Andrea~Margheri, Federico~Lombardi and Leonardo~Aniello; investigation, Sorin~Grigorescu and Federico~Lombardi; resources, Sorin~Grigorescu, Tiberiu~Cocias, Bogdan~Trasnea and Federico~Lombardi; data curation, Sorin~Grigorescu, Tiberiu~Cocias, Bogdan~Trasnea; writing--original draft preparation, Federico~Lombardi Andrea~Margheri, Leonardo~Aniello and Bogdan~Trasnea; writing--review and editing, Federico~Lombardi; visualization, Federico~Lombardi; supervision, Leonardo~Aniello; project administration, Sorin~Grigorescu; funding acquisition, Sorin~Grigorescu. All authors have read and agreed to the published version of the manuscript.
}

\funding{This research was funded by the European Commission under the CyberKit4SMEs project, grant number 883188.
}


\conflictsofinterest{The authors declare no conflict of interest.
} 


\vspace{2cm}
\abbreviations{The following abbreviations are used in this manuscript:\\

\noindent 
\begin{tabular}{@{}ll}
ADAS & Advanced Driver Assistant Systems\\
ADTF & Automotive Data and Time-Triggered Framework\\
AI & Artificial Intelligence\\
AV & Autonomous Vehicle\\
CNN & Convolutional Neural Network\\
DL & Deep Learning\\
dSPS & distributed Stream Processing Systems\\
DNN & Deep Neural Network\\
DP & Differential Privacy\\
EB & EletktroBit\\
GOL & Generative One-shot Learning\\
GPU & Graphic Processing Unit\\
HiL & Hardware-in-the-Loop\\
IoT & Internet-of-Things\\
IoU & Intersection over Union\\
ML & Machine Learning\\
MLP & Multi-Layer Perceptron\\
MPC & Multi-Party Computation\\
OTA & Over-the-Air\\
ReLu & Rectified Linear Unit\\
RM & Resource Manager\\
RNN & Recurrent Neural Network\\
ROS & Robotic Operative System\\
SiL & Software-in-the-Loop\\
TC & Training Coordinator\\
\end{tabular}}


\reftitle{References}



\externalbibliography{yes}
\bibliography{bib}

\begin{thebibliography}{-------}
\providecommand{\natexlab}[1]{#1}

\bibitem[{Grigorescu} \em{et~al.}(2019){Grigorescu}, {Trasnea}, {Cocias}, and
  {Macesanu}]{Grigorescu_JFR_2020}
{Grigorescu}, S.; {Trasnea}, B.; {Cocias}, T.; {Macesanu}, G.
\newblock A Survey of Deep Learning Techniques for Autonomous Driving.
\newblock {\em Journal of Field Robotics} {\bf 2019}, {\em 37},~362--386.

\bibitem[{Villalonga} \em{et~al.}(2020){Villalonga}, {Beruvides}, {Castaño},
  and {Haber}]{Villalonga_2020}
{Villalonga}, A.; {Beruvides}, G.; {Castaño}, F.; {Haber}, R.E.
\newblock Cloud-Based Industrial Cyber–Physical System for Data-Driven
  Reasoning: A Review and Use Case on an Industry 4.0 Pilot Line.
\newblock {\em IEEE Transactions on Industrial Informatics} {\bf 2020}, {\em
  16},~5975--5984.

\bibitem[MATLAB(2018)]{Matlab2018}
MATLAB.
\newblock {\em 9.7.0.1190202 (R2019b)}; The MathWorks Inc.: Natick,
  Massachusetts,  2018.

\bibitem[AVS()]{AVS2020}
Deep Learning on AWS.
\newblock Accessed: 2020-08-12.

\bibitem[AIO()]{AIONE2020}
The Analyst Toolbox.
\newblock Accessed: 2020-08-12.

\bibitem[Salay \em{et~al.}(2017)Salay, Queiroz, and Czarnecki]{Salay2017}
Salay, R.; Queiroz, R.; Czarnecki, K.
\newblock An Analysis of {ISO} 26262: Using Machine Learning Safely in
  Automotive Software.
\newblock {\em CoRR} {\bf 2017}, {\em abs/1709.02435},
  \href{http://xxx.lanl.gov/abs/1709.02435}{{\normalfont [1709.02435]}}.

\bibitem[Abadi \em{et~al.}(2016)Abadi et~al.]{abadi2016tensorflow}
Abadi, M.; others.
\newblock TensorFlow: Large-Scale Machine Learning on Heterogeneous Distributed
  Systems.
\newblock  12th Symposium on Operating Systems Design and Implementation,
  2016, pp. 265--283.

\bibitem[Paszke \em{et~al.}(2017)Paszke, Gross, Chintala, Chanan, Yang, DeVito,
  Lin, Desmaison, Antiga, and Lerer]{paszke2017automatic}
Paszke, A.; Gross, S.; Chintala, S.; Chanan, G.; Yang, E.; DeVito, Z.; Lin, Z.;
  Desmaison, A.; Antiga, L.; Lerer, A.
\newblock Automatic differentiation in PyTorch {\bf 2017}.

\bibitem[Jia \em{et~al.}(2014)Jia et~al.]{Caffe}
Jia, Y.; others.
\newblock Caffe: Convolutional Architecture for Fast Feature Embedding.
\newblock  Proceedings of the 22Nd ACM International Conference on Multimedia;
  ACM: New York, NY, USA,  2014; pp. 675--678.
\newblock
  doi:{\changeurlcolor{black}\href{https://doi.org/10.1145/2647868.2654889}{\detokenize{10.1145/2647868.2654889}}}.

\bibitem[Batiz-Benet \em{et~al.}(2012)Batiz-Benet, Slack, Sparks, and
  Yahya]{batiz2012parallelizing}
Batiz-Benet, J.; Slack, Q.; Sparks, M.; Yahya, A.
\newblock Parallelizing machine learning algorithms.
\newblock  Proceedings of the 24th ACM Symposium on Parallelism in Algorithms
  and Architectures, Pittsburgh, PA, USA,  2012, pp. 25--27.

\bibitem[Nangare(2018)]{Nangare2018}
Nangare, S.
\newblock Gartner's strategic tech trends show the need for an empowered edge
  and network for a smarter world,  2018.

\bibitem[Wiles(2018)]{Wiles2018}
Wiles, J.
\newblock Top Risks for Legal and Compliance Leaders in 2018,  2018.

\bibitem[Greenough(2016)]{greenough2016connected}
Greenough, J.
\newblock THE CONNECTED CAR REPORT: Forecasts, competing technologies, and
  leading manufacturers.
\newblock {\em Business Insider} {\bf 2016}.

\bibitem[Khurram \em{et~al.}(2016)Khurram et~al.]{khurram2016enhancing}
Khurram, M.; others.
\newblock Enhancing connected car adoption: Security and over the air update
  framework.
\newblock  2016 IEEE 3rd World Forum on Internet of Things. IEEE,  2016, pp.
  194--198.

\bibitem[Huang \em{et~al.}(2017)Huang, Ma, Fan, Liu, and Gong]{HuangMFLG17}
Huang, Y.; Ma, X.; Fan, X.; Liu, J.; Gong, W.
\newblock When deep learning meets edge computing.
\newblock  {ICNP}. {IEEE} Computer Society,  2017, pp. 1--2.

\bibitem[{Li} \em{et~al.}(2018){Li}, {Ota}, and {Dong}]{8270639}
{Li}, H.; {Ota}, K.; {Dong}, M.
\newblock Learning IoT in Edge: Deep Learning for the Internet of Things with
  Edge Computing.
\newblock {\em IEEE Network} {\bf 2018}, {\em 32},~96--101.
\newblock
  doi:{\changeurlcolor{black}\href{https://doi.org/10.1109/MNET.2018.1700202}{\detokenize{10.1109/MNET.2018.1700202}}}.

\bibitem[{Huang} \em{et~al.}(2018){Huang} et~al.]{8487352}
{Huang}, Y.; others.
\newblock Task Scheduling with Optimized Transmission Time in Collaborative
  Cloud-Edge Learning.
\newblock  2018 27th International Conference on Computer Communication and
  Networks (ICCCN),  2018, pp. 1--9.
\newblock
  doi:{\changeurlcolor{black}\href{https://doi.org/10.1109/ICCCN.2018.8487352}{\detokenize{10.1109/ICCCN.2018.8487352}}}.

\bibitem[Bengio \em{et~al.}(2013)Bengio, Courville, and
  Vincent]{bengio2013representation}
Bengio, Y.; Courville, A.; Vincent, P.
\newblock Representation learning: A review and new perspectives.
\newblock {\em IEEE transactions on pattern analysis and machine intelligence}
  {\bf 2013}, {\em 35},~1798--1828.

\bibitem[Hinton and Zemel(1994)]{hinton1994autoencoders}
Hinton, G.E.; Zemel, R.S.
\newblock Autoencoders, minimum description length and Helmholtz free energy.
\newblock  Advances in neural information processing systems,  1994, pp. 3--10.

\bibitem[Srivastava \em{et~al.}(2014)Srivastava, Hinton, Krizhevsky, Sutskever,
  and Salakhutdinov]{srivastava2014dropout}
Srivastava, N.; Hinton, G.; Krizhevsky, A.; Sutskever, I.; Salakhutdinov, R.
\newblock Dropout: A simple way to prevent neural networks from overfitting.
\newblock {\em The Journal of Machine Learning Research} {\bf 2014}, {\em
  15},~1929--1958.

\bibitem[Russakovsky \em{et~al.}(2015)Russakovsky, Deng, Su, Krause, Satheesh,
  Ma, Huang, Karpathy, Khosla, Bernstein, Berg, and Fei-Fei]{ImageNet_ILSVRC15}
Russakovsky, O.; Deng, J.; Su, H.; Krause, J.; Satheesh, S.; Ma, S.; Huang, Z.;
  Karpathy, A.; Khosla, A.; Bernstein, M.; Berg, A.C.; Fei-Fei, L.
\newblock {ImageNet Large Scale Visual Recognition Challenge}.
\newblock {\em International Journal of Computer Vision (IJCV)} {\bf 2015},
  {\em 115},~211--252.
\newblock
  doi:{\changeurlcolor{black}\href{https://doi.org/10.1007/s11263-015-0816-y}{\detokenize{10.1007/s11263-015-0816-y}}}.

\bibitem[Geiger \em{et~al.}(2013)Geiger, Lenz, Stiller, and Urtasun]{KITTI2013}
Geiger, A.; Lenz, P.; Stiller, C.; Urtasun, R.
\newblock {Vision Meets Robotics: The KITTI Dataset}.
\newblock {\em The Int. Journal of Robotics Research} {\bf 2013}, {\em
  32},~1231--1237.

\bibitem[Caesar \em{et~al.}(2019)Caesar, Bankiti, Lang, Vora, Liong, Xu,
  Krishnan, Pan, Baldan, and Beijbom]{nuscenes2019}
Caesar, H.; Bankiti, V.; Lang, A.H.; Vora, S.; Liong, V.E.; Xu, Q.; Krishnan,
  A.; Pan, Y.; Baldan, G.; Beijbom, O.
\newblock {NuScenes: A multimodal Dataset for Autonomous Driving}.
\newblock {\em arXiv preprint arXiv:1903.11027} {\bf 2019}.

\bibitem[Cityscapes(2018)]{Cityscapes2018}
Cityscapes.
\newblock {Cityscapes Data Collection}.
\newblock \url{https://www.cityscapes-dataset.com/},  2018.

\bibitem[Grigorescu(2018)]{grigorescu2018generative}
Grigorescu, S.M.
\newblock Generative One-Shot Learning (GOL): A Semi-Parametric Approach to
  One-Shot Learning in Autonomous Vision.
\newblock  2018 IEEE International Conference on Robotics and Automation
  (ICRA). IEEE,  2018, pp. 7127--7134.

\bibitem[Hall \em{et~al.}(2009)Hall, Frank, Holmes, Pfahringer, Reutemann, and
  Witten]{hall2009weka}
Hall, M.; Frank, E.; Holmes, G.; Pfahringer, B.; Reutemann, P.; Witten, I.H.
\newblock {The WEKA data mining software: an update}.
\newblock {\em ACM SIGKDD explorations newsletter} {\bf 2009}, {\em
  11},~10--18.

\bibitem[Xing \em{et~al.}(2016)Xing, Ho, Xie, and Wei]{xing2016strategies}
Xing, E.P.; Ho, Q.; Xie, P.; Wei, D.
\newblock Strategies and principles of distributed machine learning on big
  data.
\newblock {\em Engineering} {\bf 2016}, {\em 2},~179--195.

\bibitem[Toshniwal \em{et~al.}(2014)Toshniwal, Taneja, Shukla, Ramasamy, Patel,
  Kulkarni, Jackson, Gade, Fu, Donham, et~al.]{toshniwal2014storm}
Toshniwal, A.; Taneja, S.; Shukla, A.; Ramasamy, K.; Patel, J.M.; Kulkarni, S.;
  Jackson, J.; Gade, K.; Fu, M.; Donham, J.; others.
\newblock Storm@ twitter.
\newblock  Proceedings of the 2014 ACM SIGMOD international conference on
  Management of data. ACM,  2014, pp. 147--156.

\bibitem[Zaharia \em{et~al.}(2010)Zaharia, Chowdhury, Franklin, Shenker, and
  Stoica]{zaharia2010spark}
Zaharia, M.; Chowdhury, M.; Franklin, M.J.; Shenker, S.; Stoica, I.
\newblock Spark: Cluster computing with working sets.
\newblock {\em HotCloud} {\bf 2010}, {\em 10},~95.

\bibitem[Noel \em{et~al.}(2016)Noel, Shi, and Feng]{noel2016large}
Noel, C.; Shi, J.; Feng, A.
\newblock Large scale distributed deep learning on hadoop clusters,  2016.

\bibitem[Arimo(2016)]{tensorspark}
Arimo.
\newblock Distributed TensorFlow: Scaling Google’s Deep Learning Library on
  Spark,  2016.

\bibitem[Yang \em{et~al.}(2019)Yang, Liu, Chen, and Tong]{YangLCT19}
Yang, Q.; Liu, Y.; Chen, T.; Tong, Y.
\newblock Federated Machine Learning: Concept and Applications.
\newblock {\em {ACM} {TIST}} {\bf 2019}, {\em 10},~12:1--12:19.
\newblock
  doi:{\changeurlcolor{black}\href{https://doi.org/10.1145/3298981}{\detokenize{10.1145/3298981}}}.

\bibitem[Konecn{\'{y}} \em{et~al.}(2016)Konecn{\'{y}}, McMahan, Ramage, and
  Richt{\'{a}}rik]{KonecnyMRR16}
Konecn{\'{y}}, J.; McMahan, H.B.; Ramage, D.; Richt{\'{a}}rik, P.
\newblock Federated Optimization: Distributed Machine Learning for On-Device
  Intelligence.
\newblock {\em CoRR} {\bf 2016}, {\em abs/1610.02527},
  \href{http://xxx.lanl.gov/abs/1610.02527}{{\normalfont [1610.02527]}}.

\bibitem[{Chen} \em{et~al.}(2019){Chen}, {Li}, {Deng}, {Li}, and {Yu}]{8681645}
{Chen}, J.; {Li}, K.; {Deng}, Q.; {Li}, K.; {Yu}, P.S.
\newblock Distributed Deep Learning Model for Intelligent Video Surveillance
  Systems with Edge Computing.
\newblock {\em IEEE Transactions on Industrial Informatics} {\bf 2019}, pp.
  1--1.
\newblock
  doi:{\changeurlcolor{black}\href{https://doi.org/10.1109/TII.2019.2909473}{\detokenize{10.1109/TII.2019.2909473}}}.

\bibitem[Luckow \em{et~al.}(2016)Luckow, Cook, Ashcraft, Weill, Djerekarov, and
  Vorster]{luckow2016deep}
Luckow, A.; Cook, M.; Ashcraft, N.; Weill, E.; Djerekarov, E.; Vorster, B.
\newblock Deep learning in the automotive industry: Applications and tools.
\newblock  2016 IEEE International Conference on Big Data (Big Data). IEEE,
  2016, pp. 3759--3768.

\bibitem[Brilli \em{et~al.}(2018)Brilli, Burgio, and
  Bertogna]{brilli2018convolutional}
Brilli, G.; Burgio, P.; Bertogna, M.
\newblock Convolutional Neural Networks on embedded automotive platforms: a
  qualitative comparison.
\newblock  2018 International Conference on High Performance Computing \&
  Simulation (HPCS). IEEE,  2018, pp. 496--499.

\bibitem[Fridman \em{et~al.}(2017)Fridman et~al.]{fridman2017autonomous}
Fridman, L.; others.
\newblock Mit autonomous vehicle technology study: Large-scale deep learning
  based analysis of driver behavior and interaction with automation.
\newblock {\em arXiv preprint arXiv:1711.06976} {\bf 2017}.

\bibitem[of~Autonomous~Engineers(2019)]{SAE}
of~Autonomous~Engineers, S.
\newblock {SAE J3016 - Levels of Driving Automation},  2019.

\bibitem[Litman(2019)]{litman2019autonomous}
Litman, T.
\newblock {\em Autonomous vehicle implementation predictions}; Victoria
  Transport Policy Institute Victoria, Canada,  2019.

\bibitem[Lu \em{et~al.}(2019)Lu, Yao, and Shi]{234807}
Lu, S.; Yao, Y.; Shi, W.
\newblock Collaborative Learning on the Edges: A Case Study on Connected
  Vehicles.
\newblock  2nd {USENIX} Workshop on Hot Topics in Edge Computing (HotEdge 19);
  {USENIX} Association: Renton, WA,  2019.

\bibitem[Jiang \em{et~al.}(2019)Jiang, Lou, Tan, and Zhao]{3324405}
Jiang, L.; Lou, X.; Tan, R.; Zhao, J.
\newblock Differentially Private Collaborative Learning for the IoT Edge.
\newblock  Proceedings of the 2019 International Conference on Embedded
  Wireless Systems and Networks; Junction Publishing: USA,  2019; EWSN '19, pp.
  341--346.

\bibitem[{Yuan} and {Yu}(2014)]{6410315}
{Yuan}, J.; {Yu}, S.
\newblock Privacy Preserving Back-Propagation Neural Network Learning Made
  Practical with Cloud Computing.
\newblock {\em IEEE Transactions on Parallel and Distributed Systems} {\bf
  2014}, {\em 25},~212--221.
\newblock
  doi:{\changeurlcolor{black}\href{https://doi.org/10.1109/TPDS.2013.18}{\detokenize{10.1109/TPDS.2013.18}}}.

\bibitem[{Mohassel} and {Zhang}(2017)]{7958569}
{Mohassel}, P.; {Zhang}, Y.
\newblock SecureML: A System for Scalable Privacy-Preserving Machine Learning.
\newblock  2017 IEEE Symposium on Security and Privacy (SP),  2017, pp. 19--38.
\newblock
  doi:{\changeurlcolor{black}\href{https://doi.org/10.1109/SP.2017.12}{\detokenize{10.1109/SP.2017.12}}}.

\bibitem[Bogdanov \em{et~al.}(2008)Bogdanov, Laur, and
  Willemson]{Bogdanov:2008}
Bogdanov, D.; Laur, S.; Willemson, J.
\newblock Sharemind: A Framework for Fast Privacy-Preserving Computations.
\newblock  Proceedings of the 13th European Symposium on Research in Computer
  Security: Computer Security; Springer-Verlag: Berlin, Heidelberg,  2008;
  ESORICS '08, pp. 192--206.
\newblock
  doi:{\changeurlcolor{black}\href{https://doi.org/10.1007/978-3-540-88313-5-13}{\detokenize{10.1007/978-3-540-88313-5-13}}}.

\bibitem[Abadi \em{et~al.}(2016)Abadi, Chu, Goodfellow, McMahan, Mironov,
  Talwar, and Zhang]{abadi2016deep}
Abadi, M.; Chu, A.; Goodfellow, I.; McMahan, H.B.; Mironov, I.; Talwar, K.;
  Zhang, L.
\newblock Deep learning with differential privacy.
\newblock  Proceedings of the 2016 ACM SIGSAC Conference on Computer and
  Communications Security,  2016, pp. 308--318.

\bibitem[Lombardi \em{et~al.}(2017)Lombardi, Aniello, Bonomi, and
  Querzoni]{lombardi2017elastic}
Lombardi, F.; Aniello, L.; Bonomi, S.; Querzoni, L.
\newblock Elastic symbiotic scaling of operators and resources in stream
  processing systems.
\newblock {\em IEEE Transactions on Parallel and Distributed Systems} {\bf
  2017}, {\em 29},~572--585.

\bibitem[Badrinarayanan \em{et~al.}(2017)Badrinarayanan, Kendall, and
  Cipolla]{badrinarayanan2015segnet}
Badrinarayanan, V.; Kendall, A.; Cipolla, R.
\newblock SegNet: A Deep Convolutional Encoder-Decoder Architecture for Image
  Segmentation.
\newblock {\em IEEE Transactions on Pattern Analysis and Machine Intelligence}
  {\bf 2017}.

\bibitem[{Redmon} \em{et~al.}(2016){Redmon}, {Divvala}, {Girshick}, and
  {Farhadi}]{RedmonYOLO2016}
{Redmon}, J.; {Divvala}, S.; {Girshick}, R.; {Farhadi}, A.
\newblock You Only Look Once: Unified, Real-Time Object Detection.
\newblock  2016 IEEE Conference on Computer Vision and Pattern Recognition
  (CVPR),  2016, pp. 779--788.
\newblock
  doi:{\changeurlcolor{black}\href{https://doi.org/10.1109/CVPR.2016.91}{\detokenize{10.1109/CVPR.2016.91}}}.

\bibitem[Marina \em{et~al.}(2019)Marina, Trasnea, Tiberiu, Vasilcoi,
  Moldoveanu, and Grigorescu]{marina2019dgn}
Marina, L.; Trasnea, B.; Tiberiu, C.; Vasilcoi, A.; Moldoveanu, F.; Grigorescu,
  S.
\newblock Deep Grid Net (DGN): A Deep Learning System for Real-Time Driving
  Context Understanding.
\newblock  Int. Conf. on Robotic Computing IRC 2019; ,  2019.

\bibitem[Cam()]{CamVid2018}
The Cambridge-driving Labeled Video Database.
\newblock Accessed: 2019-07-23.

\end{thebibliography}





\end{document}